\documentclass[12pt]{article}
\usepackage{a4wide}
\usepackage{amssymb}
\usepackage{graphicx}
\begin{document}
{\renewcommand{\thefootnote}{\fnsymbol{footnote}}
\hfill  IGPG--06/11--4, AEI--2006--086\\
\medskip
\hfill gr--qc/0611112\\
\medskip
\begin{center}
{\LARGE  Effective constraints of loop quantum gravity}\\
\vspace{1.5em}
Martin Bojowald$^1$\footnote{e-mail address: {\tt
bojowald@gravity.psu.edu}}, Hector H.~Hern\'andez$^2$\footnote{e-mail
address: {\tt hehe@aei.mpg.de}}, Mikhail Kagan$^1$\footnote{e-mail
address: {\tt mak411@psu.edu}}
and Aureliano
Skirzewski$^2$\footnote{e-mail address: {\tt skirz@aei.mpg.de}}
\\
\vspace{0.5em}
$^1$Institute for Gravitational Physics and Geometry,
The Pennsylvania State
University,\\
104 Davey Lab, University Park, PA 16802, USA\\
\vspace{0.5em}
$^2$Max-Planck-Institut f\"ur Gravitationsphysik, Albert-Einstein-Institut,\\
Am M\"uhlenberg 1, D-14476 Potsdam, Germany\vspace{1.5em}
\end{center}
}

\setcounter{footnote}{0}

\newcommand{\vertex}{v}

\newcommand{\case}[2]{{\textstyle \frac{#1}{#2}}}
\newcommand{\lP}{\ell_{\mathrm P}}

\newcommand{\md}{{\mathrm{d}}}
\newcommand{\tr}{\mathop{\mathrm{tr}}\nolimits}
\newcommand{\sgn}{\mathop{\mathrm{sgn}}}

\newcommand{\id}{{\rm 1\:\!\!\! I}}

\newcommand*{\R}{{\mathbb R}}
\newcommand*{\N}{{\mathbb N}}
\newcommand*{\Z}{{\mathbb Z}}
\newcommand*{\Q}{{\mathbb Q}}
\newcommand*{\C}{{\mathbb C}}
\newcommand{\be}{\begin{equation}}
\newcommand{\ee}{\end{equation}}
\newcommand{\bq}{\begin{eqnarray}}
\newcommand{\eq}{\end{eqnarray}}

\def\f{\frac}
\def\t{\tilde}
\def\H{{\cal H}}
\newcommand{\LT}{20}
\newcommand{\LL}{10}

\begin{abstract}
 Within a perturbative cosmological regime of loop quantum gravity
 corrections to effective constraints are computed. This takes into
 account all inhomogeneous degrees of freedom relevant for scalar
 metric modes around flat space and results in explicit expressions
 for modified coefficients and of higher order terms. It also
 illustrates the role of different scales determining the relative
 magnitude of corrections. Our results demonstrate that loop quantum
gravity has the correct classical limit, at least in its sector of
cosmological perturbations around flat space, in the sense of
perturbative effective theory.
\end{abstract}

\section{Introduction}

Interacting quantum theories in low energy or semiclassical regimes
can be described by effective equations which amend the classical ones
by correction terms. Compared to the full quantum description an
analysis of such effective equations is much easier once they have
been derived. In addition to the simpler mathematical structure,
difficult conceptual and interpretational problems of wave functions
can be evaded, still allowing one to compute potentially observable
effects. Technical and conceptual problems are even more severe in
quantum gravity, in particular in background independent
formulations. Yet, especially in this case observational guidance
would be of invaluable help. Since the high energy regimes of
cosmology are commonly considered as the best access to such guidance,
an effective description for fields relevant for early universe
cosmology is needed. In this paper we use an effective framework of
perturbative loop quantum gravity around a spatially flat isotropic
background space-time to derive correction terms to the classical
constraints.

Our analysis will be done for scalar metric modes in longitudinal
gauge as this can be used to simplify the perturbative basic
variables. They can then be chosen to be of diagonal form, although
now fully inhomogeneous, which is the main reason for simplifications
as they have been used extensively in symmetric models
\cite{IsoCosmo,HomCosmo,SphSymm,LivRev}. The main constructions of
these models can thus be extended, in a similarly explicit form, to
inhomogeneous situations without assuming any symmetry. This allows us
to compute explicit corrections to effective constraints which, in
combination with the Hamiltonian analysis of cosmological perturbation
theory in \cite{HamPerturb}, leads to corrected perturbation equations
and new effects \cite{InhomEvolve}.

A physical regime is selected by introducing a background geometry in
the background independent quantization through a specific class of
states \cite{InhomLattice}. This keeps the characteristic background
independent features of the quantum theory, such as its spatial
discreteness, intact while bringing the theory to a form suitable for
perturbative expansions and applications.  In the perturbative regime,
we will make use of special structures provided by the geometrical
background which can usually be chosen to allow symmetries, e.g.\
isotropy for cosmological perturbations around a
Friedmann--Robertson--Walker model. In particular, we use this to
introduce regular lattice states with a spacing (in geometrical
variables rather than embedding coordinates) whose size determines at
which scales quantum effects become important. The geometrical spacing
thus specifies on which scales physical fields are being probed by a
given class of states. On such a lattice, explicit calculations can be
done.

We demonstrate this by providing higher curvature terms as well as
corrections to inverse powers of metric components. Several issues
that arose in isotropic models will be clarified. Finally, we discuss
general aspects of effective equations and the semiclassical limit of
loop quantum gravity. The article thus consists of two parts, an
explicit scheme to derive correction terms presented in
Sec.~\ref{s:HamConstr} and \ref{EffHamDiscuss}, and a more general
discussion of effective equations and the classical limit in
Sec.~\ref{Effective}.

\section{Basic variables and operators}

The basic variables of interest for a canonical formulation of gravity
\cite{ADM} are the spatial metric $q_{ab}$ occurring in the space-time
metric
\begin{equation} \label{metric}
 \md s^2=-N^2\md t^2+q_{ab}(\md x^a+N^a\md t) (\md x^b+N^b\md t)\,.
\end{equation}
(or equivalent quantities such as a triad $e^i_a$ with
$e^i_ae^i_b=q_{ab}$ or its inverse $e^a_i$), extrinsic curvature
$K_{ab}=(2N)^{-1}(\dot{q}_{ab}-D_aN_b-D_bN_a)$ (or related objects
such as the Ashtekar connection) and matter fields with their
momenta. The components $N$ (lapse function) and $N^{a}$ (shift
vector) of the space-time metric are not dynamical, and thus do not
have momenta, but are important for selecting the space-time slicing
or the gauge.

Because of their role in loop quantum gravity, we will use Ashtekar
variables \cite{AshVar,AshVarReell} which are a densitized triad
$E^a_i=|\det e^j_b| e^a_i$ and the connection $A_a^i=\Gamma_a^i-\gamma
K_a^i$ with the spin connection
\begin{equation} \label{SpinConnFull}
 \Gamma_a^i= -\epsilon^{ijk}e^b_j (\partial_{[a}e_{b]}^k+
 {\textstyle\frac{1}{2}} e_k^ce_a^l\partial_{[c}e_{b]}^l)\,.
\end{equation}
compatible with the triad, $K_a^i=K_{ab}e^b_i$ and the positive
Barbero--Immirzi parameter $\gamma$ \cite{AshVarReell,Immirzi}.  We
use them here in perturbative form on a flat isotropic metric
background, focusing on scalar modes. This means, as explained in more
detail in \cite{HamPerturb}, that the unperturbed metric as well as
its perturbations can be assumed to be diagonal,
$E^a_i=\tilde{p}^{(i)}(x)\delta^a_i$, which simplifies
calculations. For scalar modes, all diagonal components of the metric
$q_{ab}=a^2(1-2\psi(x)^2)\delta_{ab}$ are in fact equal, but we will
see that this restriction is not general enough for formulating a loop
quantization. Moreover, we can choose a vanishing shift vector
$N^a=0$, implying that extrinsic curvature
$K_a^i=\tilde{k}_{(i)}(x)\delta_a^i$ is diagonal, too. (The Ashtekar
connection, on the other hand, will not be diagonal because it has
non-diagonal contributions from the spin connection. It is of the form
$A_a^i=\tilde{k}_{(i)}(x)\delta_a^i+\psi_I(x)\epsilon^{iI}_a$ where
the non-diagonal part $\psi_I$ arising from the spin connection
computed in Sec.~\ref{SpinConn} can be
dealt with perturbatively.)  Our calculations will thus be done in a
given gauge, and would be more complicated in others. Nevertheless, we
are including the general perturbations of metric and matter variables
relevant for scalar modes without too strong restrictions as they
could arise in other gauges.

\subsection{Gauge choices and their implications for a quantization}

In general, the space-time gauge is determined by prescribing the
behavior of lapse function $N$ and shift vector $N^a$ occurring in a
metric (\ref{metric}).  The lapse function, as we will see, can be
chosen arbitrarily in our calculations, but the shift vector is
restricted for a diagonal perturbation to be realized. We are thus
using a particular class of gauges in setting up our calculations,
although we do not explicitly make use of the form of gauge
transformations. This is important because the canonical constraints,
most importantly the Hamiltonian constraint $H$ in addition to the
diffeomorphism constraint $D_a$, and thus also the gauge
transformations $\delta_{\xi}f=\{f,\xi^0H+\xi^aD_a\}$ they generate
will be corrected by quantum effects. Classical properties of the
gauge transformations should thus not be used before one computes
quantum corrections. It is then a priori unclear which particular
gauge choices, other than fixing lapse and shift directly, are
allowed. Some gauges implicitly refer to gauge transformation
equations to relate metric to matter perturbations, or to select the
space-time slicing such as for the flat gauge. In this example, one
would make use of gauge transformations to set the spatial metric
perturbation equal to zero which allows one to focus calculations on
the simpler matter part. In this process, one solves gauge
transformation equations of the metric perturbation, depending on
lapse and shift, such that the transformed perturbation vanishes. This
determines a gauge to be chosen, but makes use of explicit gauge
transformation equations which are not guaranteed to remain unchanged
with quantum corrections. Our choice of vanishing shift, on the other
hand, is harmless because it does not refer to explicit gauge
transformation equations. We are thus working at a more general level
keeping metric and matter perturbations independent. A combined gauge
invariant combination of the two perturbations can be determined once
the quantum corrected gauge transformations have been computed.

When constraints are modified, manifest covariance of the resulting
equation becomes an issue as it is discussed in more detail in
\cite{HamPerturb}. Such quantum corrections are derived from
effective constraints of gravity which are defined as expectation
values of quantum gravity operators in general states \cite{EffAc}. We
motivate the procedure here briefly, and will provide some further
information in Sec.~\ref{Effective}; for details we refer to
\cite{EffAc,Karpacz,EffectiveEOM}. If constraint operators satisfy the
classical constraint algebra, covariance would be manifest for the
expectation values. But there is an additional step involved in
deriving effective equations: the expectation values depend on
infinitely many quantum variables such as the spreads of states which
do not have classical analogs. Effective equations are obtained by
truncating this to finitely many fields (similarly to the derivative
expansion in low energy effective actions), resulting in equations of
motion of the classical form corrected by quantum terms.  Indeed, any
quantum theory is based on states which are not just determined by
expectation values of the basic variables such as $A_a^i$ and $E^a_i$
in loop quantum gravity.  Expectation values of the basic variables
would correspond to the classical values in constraint expressions,
but there are infinitely many further parameters such as the spread
and deformations of the state from a Gaussian. These additional,
quantum variables can suitably be parameterized in the form
\begin{equation} \label{QuantVars}
 G^{a,n}_q=\langle(\hat{q}-\langle\hat{q}\rangle)^{n-a}
(\hat{p}_q-\langle\hat{p}_q\rangle)^a\rangle_{\rm Weyl}
\end{equation}
for any degree of freedom $(q,p_q)$ present in the classical
theory. Here, $1<n\in{\mathbb N}$, $0\leq a\leq n$, and the subscript
``Weyl'' denotes totally symmetric ordering. Every classical degree of
freedom thus does not only give rise to expectation values
$\langle\hat{q}\rangle$ and $\langle\hat{p}_q\rangle$ but to infinitely
many additional quantum variables. All of these variables are
dynamical, and are in general coupled to each other.

Moreover, expectation values of most operators, including
Hamiltonians, in general states depend on all these infinitely many
variables. This is to be reduced to a finite set for an effective
description which introduces quantum correction terms into the
classical equations. In particular, spread and deformation parameters
are usually assumed to be subdominant compared to expectation
values. Without explicitly constructing semiclassical states
satisfying such conditions, one can make semiclassicality assumptions
for those parameters to be negligible. This is what we will do in this
paper as a shortcut to deriving effective expressions from a full
quantum theory. Since special quantization steps are involved in the
construction of operators which reformulate classical expressions,
corrections in effective equations will result which are not sensitive
to the precise form of semiclassical states.

\subsection{Lattice states and basic operators}

We are thus able to implement all degrees of freedom needed for
inhomogeneities in a way which is accessible to explicit
calculations. While the general kinematical arena of loop quantum
gravity is based on discrete spatial structures built on arbitrary
graphs with possibly high-valent vertices, we will use regular
lattices with 6-valent vertices. Regularity of the lattice is
implemented by making use of symmetries of the background we are
perturbing around: The three independent generators of translational
symmetry define lattice directions. In explicit constructions of
lattice states, a scale $\ell_0$ will appear which is the coordinate
length of lattice links measured in a given, fixed
embedding.\footnote{The coordinate length need not be the same for all
links, but can be chosen this way without loss of generality.} But
this parameter is independent of the quantum variables assigned to
each link we will be using, which means that the quantum theory will
be defined on ``freely floating'' lattices as in the full theory,
respecting diffeomorphism invariance. The scale $\ell_0$ will only
become important in the continuum limit, when making contact between
the quantum variables and classical continuous fields. This obviously
breaks manifest diffeomorphism covariance, just as the classical
perturbation theory in basic fields rather than gauge-invariant
combinations, since the classical perturbations are written with
respect to a background space-time.

Compared to the full theory, we are restricting states by assuming
regularity and thus allowing, e.g., only unknotted links and vertices
of valence at most six. This turns out to be sufficient to include all
relevant perturbative degrees of freedom. While the general graphs of
loop quantum gravity allow more freedom, its meaning is not known and
appears redundant in our application.

\subsubsection{Holonomies and fluxes}

The canonical fields are given by $(A_a^i,E^b_j)$ which are to be
turned into operators on a suitable Hilbert space. To set this up, we
need to choose a functional representation of state, which is
conveniently done in the connection representation where states are
functionals of $A_a^i$.  According to loop quantum gravity, lattice
graphs then label states and determine their expressions as functions
of connections: a state associated with a given graph depends on the
connection only through holonomies
\[
 h_e(A)={\cal P}\exp(\smallint_e\md t A^i_a\dot{e}^a\tau_i)
\]
along its edges. Here $\tau_j=-\f{i}{2}\sigma_j$ are the
$SU(2)$-generators in terms of Pauli matrices $\sigma_j$ and ${\cal
P}$ denotes path ordering. That those are the basic objects
represented on a Hilbert space together with fluxes
\begin{equation}\label{flux}
 F_S(E)=\int_S\md^2y E^a_i\tau^in_a
\end{equation}
for surfaces $S$ with co-normal $n_a$ is the basic assumption of loop
quantum gravity \cite{LoopRep}. Our corrections to cosmological
perturbation equations will be implications of this fact and thus test
the theory directly. Using the perturbative form of $A_a^i$, we can
split off perturbatively the non-diagonal part (composed of spin
connection components) in an expansion and exploit the diagonality of
the remaining part to obtain $h_{v,I}=\exp
(\gamma\tau_I\int_{e_{v,I}}\md t \tilde{k}_I(e_{v,I}(t))$. Similarly,
fluxes will be of the form $F_{v,I}=\int_{S_{v,I}}\md^2y
\tilde{p}^I(y)$. A lattice link starting at a vertex $v$ in direction
$I$ in a fixed orientation is denoted $e_{v,I}$, and a lattice
plaquette transversal to this edge and centered at its midpoint as
$S_{v,I}$. (Here and in the following set-up we closely follow
\cite{InhomLattice} to which we refer the reader for more details;
note, however, that $\gamma$ has been absorbed there in
$\tilde{k}_I$.)

Matrix elements of the variables $h_{v,I}$ together with $F_{v,I}$
form the basic objects of loop quantum gravity in this setting. They
are thus elementary degrees of freedom, comparable to atoms in
condensed matter. Classical fields will, as we display in detail
later, emerge from these objects in suitable regimes and limits
only. Even in such regimes where one can recover the usual metric
perturbations there will in general be correction terms examples of
which we aim to compute below. Correction functions will then also
depend on the basic objects $h_{v,I}$ and $F_{v,I}$ directly, which
can be expressed through the classical metric perturbations in a
secondary step.\footnote{This already indicates that difficulties
which were sometimes perceived in isotropic models, where corrections
seemed to depend on the scale factor whose total scale is
undetermined, do not occur in this inhomogeneous setup.}

To recover the correct semiclassical behavior one has to make sure
that effective equations of motion can indeed be written in a form
close to the classical ones. Since classical Hamiltonians are local
functionals of extrinsic curvature and densitized triad components, it
must then be possible to approximate the non-local, integrated objects
$h_{v,I}$ and $F_{v,I}$ by local values of $\tilde{k}_I$ and
$\tilde{p}^I$ evaluated in single points This is indeed possible if we
assume that $\tilde{k}_I$ is approximately constant along any edge,
whose coordinate lengths are $\ell_0=\int_{e_{\vertex,I}}\md t$.  We
can then write
\begin{equation}\label{hol}
 h_{v,I}=\exp
 (\smallint_{e_{\vertex,I}}\md t \gamma\tilde{k}_I\tau^I)\approx
 \cos(\ell_0\gamma\tilde{k}_{I}(\vertex+I/2)/2)+
 2\tau_I\sin(\ell_0\gamma\tilde{k}_{I}(\vertex+I/2)/2)
\end{equation}
where $\vertex+I/2$ denotes, in a slight abuse of notation, the
midpoint of the edge which we use as the most symmetric relation
between holonomies and continuous fields, and
\begin{equation} \label{ScalarFlux}
 F_{\vertex,I}=
\int_{S_{\vertex,I}} \tilde{p}^I(y)\md^2y\approx
\ell_0^2\tilde{p}^I(\vertex+I/2)
\end{equation}
(note that the surface $S_{v,I}$ is defined to be centered at the
midpoint of the edge $e_{v,I}$).  This requires the lattice to be fine
enough, which will be true in regimes where fields are not strongly
varying. For more general regimes this assumption has to be dropped
and non-local objects appear even in effective approximations since a
function $\tilde{k}_I$ will be underdetermined in terms of the
$h_{v,I}$. Since the recovered classical fields must be continuous,
this means that they can arise only if quantizations of $h_{v,I}$ and
$F_{v,I}$, respectively, for nearby lattice links do not have too much
differing expectation values in a semiclassical state. If this is not
satisfied, continuous classical fields can only be recovered after a
process of coarse graining as we will briefly discuss in
Sec.~\ref{Coarse}.

In addition to the assumption of slowly varying fields on the lattice
scale, we have also made use of the diagonality of extrinsic curvature
which allows us to evaluate the holonomy in a simple way without
taking care of the factor ordering of su(2)-values along the path. We
can thus re-formulate the theory in terms of U(1)-holonomies
\begin{equation}
 \eta_{\vertex,I} = \exp(i\smallint_{e_{\vertex,I}}\md t
 \gamma\tilde{k}_I/2)\approx \exp(i\ell_0 \gamma\tilde{k}_I(\vertex+I/2)/2)
\end{equation}
along all lattice links $e_{\vertex,I}$.  On the lattice, a basis of
all possible states is then given by specifying an integer label
$\mu_{\vertex,I}$ for each edge starting at vertex $\vertex$ in
direction $I$ and defining
\begin{equation}\label{hol_action}
 \langle \tilde{k}(x)|\ldots,\mu_{\vertex,I},\ldots\rangle:= \prod_{\vertex,I}
\exp (i\mu_{\vertex,I}\smallint_{e_{\vertex,I}}\md t
\gamma\tilde{k}_I/2)
\end{equation}
as the functional form of the state
$|\ldots,\mu_{\vertex,I},\ldots\rangle$ in the $k$-representation. The
form of the states is a consequence of the representation of
holonomies. States are functions of U(1)-holonomies, and any such
function can be expanded in terms of irreducible representations which
for U(1) are just integer powers.  This would be more complicated if
we allowed all possible, also non-diagonal, curvature components as
one is doing in the full theory. In such a case, one would not be able
to reduce the original SU(2)-holonomies to simple phase factors and
more complicated multiplication rules would have to be considered. In
particular, one would have to make sure that matrix elements of
holonomies are multiplied with each other in such a way that functions
invariant under SU(2)-gauge rotations result
\cite{RS:Spinnet}. This requires additional vertex labels which we do
not need in the perturbative situation.

\begin{figure}
\centerline{\includegraphics[width=8cm,keepaspectratio]{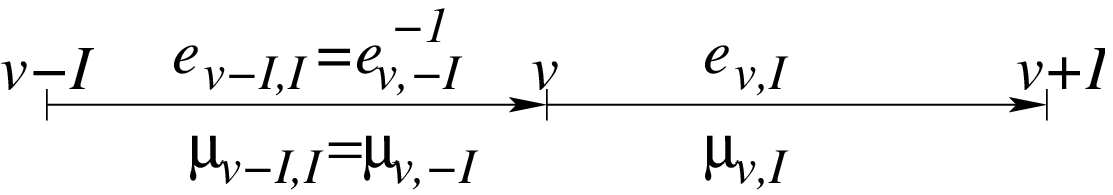}}
\caption{Edges adjacent to a vertex $v$ in a given direction $I$. For
the edge oriented oppositely to the chosen one for direction $I$, the
labels ``$v-I,I$'' and ``$v,-I$'' can be chosen interchangeably,
defining in this way negative values for the label $I$.
\label{eminusI}}
\end{figure}

For the same reason we have simple multiplication operators given by
holonomies associated with lattice links,
\begin{equation}
 \hat{\eta}_{\vertex,I}|\ldots,\mu_{\vertex',J},\ldots\rangle =
|\ldots,\mu_{\vertex,I}+1,\ldots\rangle\,.
\end{equation}
There are also derivative operators with respect to $\tilde{k}_I$,
quantizing the conjugate triad components. Just as holonomies are
obtained by integrating the connection or extrinsic curvature,
densitized triad components are integrated on surfaces, then called
fluxes (\ref{flux}), before they can be quantized. For a surface
$S$ of lattice plaquette size intersecting a single edge
$e_{\vertex,I}$ outside a vertex, we have
\begin{equation}
 \hat{F}_{\vertex,I} |\ldots,\mu_{\vertex',J},\ldots\rangle=
 4\pi\gamma\lP^2\mu_{\vertex,I}
|\ldots,\mu_{\vertex',J},\ldots\rangle
\end{equation}
or
\begin{equation} \label{FluxVert}
 \hat{{\cal F}}_{\vertex,I} |\ldots,\mu_{\vertex',J},\ldots\rangle=
 2\pi\gamma\lP^2(\mu_{\vertex-I,I}+\mu_{\vertex,I})
|\ldots,\mu_{\vertex',J},\ldots\rangle\,.
\end{equation}
if the intersection happens to be at the vertex. The Planck length
$\ell_{\rm P}=\sqrt{G\hbar}$ arises through a combination of $G$ from
the basic Poisson brackets and $\hbar$ from a quantization of momenta
as derivative operators. Here, in a similar notation as above,
$\vertex-I$ denotes the vertex preceding $\vertex$ along direction $I$
in the given orientation. We will later call such labels simply
$\mu_{v-I,I}=\mu_{v,-I}$ as illustrated in Fig.~\ref{eminusI}. These
operators quantize integrated triad components (\ref{ScalarFlux}).
This shows that all basic degrees of freedom relevant for us can be
implemented without having to use the more involved SU(2)-formulation.

Note, that even for scalar perturbations which classically have triads
proportional to the identity, distinct $\tilde{p}^I(v)$-components
have to be treated differently at the quantum level. One cannot assume
all edge labels around any given vertex to be identical while still
allowing inhomogeneity. Moreover, operators require local edge
holonomies which change one edge label $\mu_{v,I}$ independently of
the others. Similarly, corresponding operators $\hat{F}_{v,I}$ and
$\hat{F}_{v,J}$ ($I\neq J$) act on different links coming out of a
vertex $v$ and have thus independent eigenvalues in general. To pick a
regime of scalar modes, one will choose a state whose edge fluxes are
peaked close to the same triad value in all directions and whose
holonomies are peaked close to the same exponentiated extrinsic
curvature values, thus giving effective equations for a single scalar
mode function. But this to equal values in different directions cannot
be done at the level of operators.

These basic operators $h_{v,I}$ and $F_{v,I}$ will appear in more
complicated ones and in particular in the constraints. As we can see,
they depend not directly on the classical fields $\tilde{p}^I(x)$ and
$\tilde{k}_I(x)$ but, in local approximations, on quantities
$p^I(x):=\ell_0^2\tilde{p}^I(x)$ and $k_I(x):=\ell_0\tilde{k}_I(x)$
rescaled by factors of the lattice link size $\ell_0$. This re-scaling
occurring automatically in our basic variables has two advantages: It
makes the basic variables independent of coordinates and provides them
unambiguously with dimensions of length squared for $p$ while $k$
becomes dimensionless. (Otherwise, one could choose to put dimensions
in coordinates or in metric components which would make arguments for
the expected relevance of quantum corrections more complicated.) This
also happens in homogeneous models \cite{Bohr}, but in that case,
especially in spatially flat models, there was sometimes confusion
about the meaning of even the re-scaled variables. This is because the
scale factor, for instance, as the isotropic analog of $\tilde{p}^I$
could be multiplied by an arbitrary constant and thus the total scale
would have no meaning even when multiplied by the analog of
$\ell_0^2$. Thus, correction functions depending on this quantity in
an isotropic model require an additional assumption on how the total
scale is fixed.

This is not necessary in inhomogeneous situations.  Here, the
quantities $p^I$ will appear in quantum corrections and their values
determine unambiguously when corrections become important. The
corresponding fluxes are the relevant quantum excitations, and when
they are close to the Planck scale quantum corrections will
unambiguously become large. On the other hand, if the $p^I$ become too
large, approaching the Hubble length squared or a typical wave length
squared, discreteness effects become noticeable even in usual
physics. As we will see in more detail in Sec.~\ref{magnitude}, this
allows one to estimate orders of magnitudes of corrections to be
expected even without detailed calculations
\cite{InhomEvolve}. Although the size of the $p^I$ is coordinate
independent, unlike the value of the scale factor, say, its relation
to the classical field depends on $\ell_0$ and thus on the lattice
size. It may thus appear that $p^I$ is coordinate dependent, but this
is clearly not the case because it derives directly from a coordinate
independent flux. The lattice values are defined independently of
coordinates, just by attaching labels $\mu_{v,I}$ to lattice
links. Once they have been specified and the lattice has been embedded
in a spatial manifold, their relation to classical metric fields can
be determined. It is, of course, the classical fields such as metric
components which depend on the coordinate choice when they are
tensorial. The relation between $p^I$ and the classical metric depends
on the lattice spacing measured in coordinates because the
representation of the classical metric itself depends on which
coordinates have been chosen. Thus, our basic quantities are
coordinate independent and coordinates enter only when classical
descriptions are recovered in a semiclassical limit.

\subsubsection{Volume}

An important ingredient to construct constraints is the volume
operator. Using the classical expression $V=\int\md^3x
\sqrt{|\tilde{p}^1\tilde{p}^2\tilde{p}^3|}$ we introduce the
volume operator $\hat{V}=\sum_{\vertex} \prod_{I=1}^3 \sqrt{|\hat{\cal
F}_{\vertex,I}|}$ which, using (\ref{FluxVert}), has eigenvalues
\begin{equation}\label{V_action}
 V(\{\mu_{\vertex,I}\})=
\left(2\pi\gamma\lP^2\right)^{3/2}
 \sum_{\vertex} \prod_{I=1}^3\sqrt{
|\mu_{\vertex,I}+\mu_{\vertex-I,I}|}\,.
\end{equation}
While densitized triad components are directly implemented through
basic fluxes, the process of quantizing triad or co-triad components
is more indirect. While they are uniquely determined from the
densitized triad classically, one needs to take inverse
components. With flux operators having discrete spectra containing
zero, this is not possible in a direct manner at the quantum
level. Nevertheless \cite{QSDI}, one can construct operators for
co-triad components based on the classical identity
\begin{equation} \label{cotriad}
 \left\{A_a^i,\int\sqrt{|\det E|}\mathrm{d}^3x\right\}= 2\pi\gamma G
 \epsilon^{ijk}\epsilon_{abc} \frac{E^b_jE^c_k}{\sqrt{|\det E|}}
=4\pi\gamma G e_a^i\,.
\end{equation}
On the left hand side, no inverse appears and we just need to
express connection components in terms of holonomies, use the volume
operator and replace the Poisson bracket by a commutator. Resulting
operators are then of the form $h_e[h_e^{-1},\hat{V}]$ for
SU(2)-holonomies along suitable edges $e$, e.g.\
\begin{equation} \label{commPoiss}
 \tr(\tau^ih_{v,I}[h_{v,I}^{-1},\hat{V}_v])\sim -\frac{1}{2}i\hbar\ell_0
\widehat{\{A_a^i,V_v\}}
\end{equation}
for $h_{v,I}$ as in (\ref{hol}). This shows that factors of the link
size $\ell_0$ are needed in reformulating Poisson brackets through
commutators with holonomies, which, as will become clear below, are
provided by the discretized integration measure in spatial
integrations such as they occur in the Hamiltonian constraint.

\section{Hamiltonian constraint}
\label{s:HamConstr}

Holonomies, the volume operator and commutators between them are
finally used to define Hamiltonian constraint operators. We will
briefly describe the general procedure and then derive resulting
correction terms in effective equations for both gravitational and
matter contributions to the constraint.

\subsection{Gravitational part}

\begin{figure}
\centerline{\includegraphics[width=6cm,keepaspectratio]{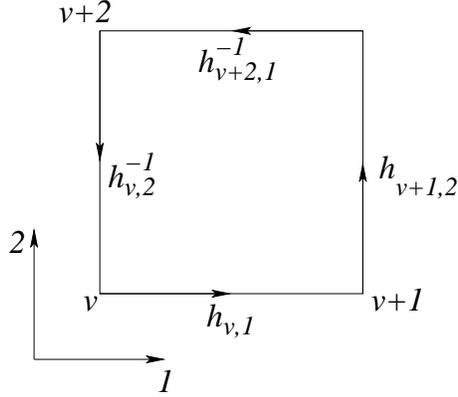}}
\caption{Elementary lattice plaquette with holonomies around a closed loop.
 \label{Plaquette}}
\end{figure}

The gravitational contribution to the Hamiltonian constraint is given
by
\begin{eqnarray}\label{HamConstr}
 H[N] &=& \frac{1}{16\pi G} \int_{\Sigma} \mathrm{d}^3x N\left|\det
   E\right|^{-1/2}
 \left(\epsilon_{ijk}F_{ab}^iE^a_jE^b_k\right.\\
 && -\left.2(1+\gamma^{-2})
 (A_a^i-\Gamma_a^i)(A_b^j-\Gamma_b^j)E^{[a}_iE^{b]}_j\right)\nonumber
\end{eqnarray}
in terms of Ashtekar variables with the curvature
$F^i_{ab}=2\partial_{[a}A^i_{b]}+\epsilon^{ijk}A_a^jA_b^k$. The second
term, quadratic in extrinsic curvature components
$K_a^i=\gamma^{-1}(\Gamma_a^i-A_a^i)$, is in general more complicated
to deal with due to the appearance of spin connection components as
functionals of $E^a_i$ through (\ref{SpinConnFull}). One usually
starts with quantizing the first term and then uses the identity
\cite{QSDI}
\begin{equation} \label{Kcomm}
 K_a^i=\gamma^{-1}(A_a^i-\Gamma_a^i)
 \propto \left\{A_a^i,\!\left\{\int\md^3x F_{ab}^i
 \frac{\epsilon^{ijk}E^a_jE^b_k}{\sqrt{|\det
 E|}},\int{\sqrt{|\det E|}}\mathrm{d}^3x\right\}\right\}
\end{equation}
which allows one to express the second contribution in terms of the
first. In the first term, then, the densitized triad components
including the inverse determinant can be quantized using
(\ref{cotriad}), and the curvature components $F_{ab}^i$ can be
obtained through holonomies around appropriately chosen small loops
\cite{RS:Ham}. On our regular lattices, natural loops based at a given
vertex are provided by the adjacent lattice plaquettes. After
replacing the Poisson brackets by commutators, the resulting first
part of the Hamiltonian operator,
$\hat{H}^{(1)}=\sum_v\hat{H}^{(1)}_v$ has non-zero action only in
vertices of a lattice state, each contribution being of the form
\begin{eqnarray} \label{Hone}
 \hat{H}_v^{(1)} &=& \frac{1}{16\pi G}\frac{2i}{8\pi\gamma
 G\hbar}\frac{N(v)}{8}
\sum_{IJK}\sum_{\sigma_I\in\{\pm1\}} \sigma_1\sigma_2\sigma_3
\epsilon^{IJK}\\
&&\times\tr(h_{\vertex,\sigma_II}(A)
h_{\vertex+\sigma_II,\sigma_JJ}(A)
h_{\vertex+\sigma_JJ,\sigma_II}(A)^{-1} h_{\vertex,\sigma_JJ}(A)^{-1}
h_{\vertex,\sigma_KK}(A) [h_{\vertex,\sigma_KK}(A)^{-1},\hat{V}])
\nonumber
\end{eqnarray}
summed over all non-planar triples of edges in all possible
orientations. (There are 48 terms in the sum, but we need to divide
only by 8 since a factor of six arises in the contraction of basic
fields occurring in the constraint.) The combination
\[
 h_{\vertex,\sigma_II}(A) h_{\vertex+\sigma_II,\sigma_JJ}(A)
 h_{\vertex+\sigma_JJ,\sigma_II}(A)^{-1} h_{\vertex,\sigma_JJ}(A)^{-1}
\]
gives a single plaquette holonomy with tangent vectors
$e_{v,\sigma_II}$ and $e_{v,\sigma_JJ}$ as illustrated in
Fig.~\ref{Plaquette}.

When expanded in $\ell_0$ assuming sufficiently small edges, the
leading term is of the order $\ell_0^3$ which automatically results in
a Riemann sum representation of the first term in
(\ref{HamConstr}). This justifies $\hat{H}^{(1)}$ as a quantization of
the classical expression. As seen from the argument, one needs to
assume that the lattice is sufficiently fine for classical values of
the fields $A_a^i$. Thus, there are states corresponding to coarser
lattices on which stronger quantum corrections can result. As usually,
semiclassical behavior is not realized on all states but only for a
select class. For any low-curvature classical configuration, one can
make sure that a chosen lattice leads only to small quantum
corrections such that sufficiently many semiclassical states exist.

\subsubsection{Quantization}

The required calculations for SU(2) holonomies and their products
usually do not allow explicit diagonalizations of operators. But some
physical regimes allow one to decouple the matrix components at least
approximately. This is realized for several symmetric models and also
for perturbations at least of some metric modes around them. In
particular, after splitting off the non-diagonal part of the
connection in the perturbative expansion considered here, we can take
the trace explicitly and reduce the expression to U(1). Since the
diagonal part of the Ashtekar connection for our perturbations is
contributed entirely by extrinsic curvature, we are effectively using
``holonomies'' computed for extrinsic curvature rather than the
Ashtekar connection. Although extrinsic curvature is a tensor rather
than a connection, it is meaningful to use it in expressions
resembling holonomies, denoted here simply as $h_{v,I}$, on a given
metric background. This has the additional advantage of easily
combining the remaining quadratic terms in $K_a^i$ with the first term
of the constraint (\ref{HamConstr}) without using squares of multiple
commutators from quantizing (\ref{Kcomm}). Writing
\bq \label{Curv}%
F_{ab}^i &=& 2\partial_{[a}\Gamma_{b]}^i +2 \gamma\partial_{[a}
K_{b]}^i+\epsilon_{ijk}\left(\Gamma_a^j+\gamma
K_a^j\right)\left(\Gamma_b^k+\gamma K_b^k\right) \nonumber \\
&=&2\partial_{[a}\Gamma_{b]}^i +2 \gamma\partial_{[a} K_{b]}^i+\gamma
\epsilon_{ijk} \left(\Gamma_a^j K_b^k+\Gamma_b^k K_a^j
\right)+\epsilon_{ijk}\left(\Gamma_a^j\Gamma_b^k+\gamma^2 K_a^j
K_b^k\right)
\eq%
we obtain a term $2 \gamma\partial_{[a} K_{b]}^i+\gamma^2 K_a^j
K_b^k$ resembling ``curvature'' $F_{ab}^i(K)$ as computed from extrinsic
curvature alone, a curvature term of the spin connection as well as
cross-terms $\epsilon_{ijk} \left(\Gamma_a^j K_b^k+\Gamma_b^k K_a^j
\right)$. In our context due to the diagonality conditions the
cross-terms disappear \cite{HamPerturb} and we only have the
$K$-curvature term and spin connection terms to quantize. The first
term can then be combined with the term quadratic in $K_a^i$ in
(\ref{HamConstr}), removing the need to use double commutators. We denote
this contribution to the constraint as
\begin{equation}
 H_K[N]:= \frac{1}{8\pi G} \int_{\Sigma} \mathrm{d}^3x N
\left|\det
   E\right|^{-1/2} \left(\epsilon_{ijk} \gamma\partial_aK_b^i
 -K_a^j K_b^k \right)E^{[a}_jE^{b]}_k
\end{equation}
(since also $\partial_aK_b^i$ drops out as used later, the constraint
is $\gamma$-independent) and the remaining term as
\begin{equation}
 H_{\Gamma}[N]:=\frac{1}{8\pi G} \int_{\Sigma} \mathrm{d}^3x N
\left|\det
   E\right|^{-1/2} \left(\epsilon_{ijk}\partial_a\Gamma_b^i+ \Gamma_a^j
\Gamma_b^k\right)E^{[a}_jE^{b]}_k\,.
\end{equation}
Then, $H[N]=H_K[N]+H_{\Gamma}[N]$ is the constraint for scalar modes
in longitudinal gauge.  Both terms can rather easily be dealt with,
using holonomies around a loop for the first term (this
subsection) and direct quantizations of $\Gamma_a^i$ for the second
(Sec.~\ref{SpinConn}). The split-off spin connection components are
thus quantized separately, which is possible in the perturbative
treatment on a background, and then added on to the operator.

Note also that as a further simplification the derivative term of
extrinsic curvature disappears from the constraint for diagonal
variables as assumed here. This will automatically happen also from
holonomies around loops. We emphasize that the quantization procedure
followed is special to the given context of scalar perturbations on a
flat isotropic background. Nevertheless, it mimics essential steps of
the full constructions as discussed in more detail in
Sec.~\ref{QuantProc}. Its main advantage is that it allows explicit
derivations of all necessary terms and thus explicit effective
equations to be confronted with observations. Moreover, it is far from
clear that the constructions currently done in the full setting will
remain unchanged with further developments. We thus evaluate the key
features of the scheme without paying too close attention to current
details.

Following the general procedure, we thus obtain vertex contributions
\begin{eqnarray}\label{Kcontrib}%
\hat{H}_{K,v}&=& -\frac{1}{16\pi G}\frac{2i}{8\pi\gamma^3 G\hbar}
\frac{N(v)}{8}
\sum_{IJK}\sum_{\sigma_I\in\{\pm1\}} \sigma_1\sigma_2\sigma_3
\epsilon^{IJK}\\
&&\times\tr \left(h_{v,\sigma_II} h_{v+\sigma_II,\sigma_JJ}
h_{v+\sigma_JJ,\sigma_II}^{-1} h_{v,\sigma_JJ}^{-1} h_{v,\sigma_KK}
\left[h_{v,\sigma_KK}^{-1},\hat
V_v\right]\right)\,.\nonumber
\end{eqnarray}
 As before, $h_{v,I}$ denotes a $K$-holonomy along
the edge oriented in the positive $I$-direction and starting at vertex
$v$, but we also include the opposite direction $h_{v,-I}$ in the sum
to ensure rotational invariance. Note that following our convention,
such holonomies are identified with $h_{v-I,I}^{-1}$.  In some of the
holonomies, $v+I$ is again the vertex adjacent to $v$ in the positive
$I$-direction. The $\{IJK\}$-summation is taken over all possible
orientations of the ${IJ}$-loop and a transversal $K$-direction. Also,
for notational brevity, we introduce
\be%
c_{v,I}:=\frac{1}{2}\tr(h_{v,I}) \quad,\quad s_{v,I}:=-\tr(\tau_{(I)}h_{v,I})
\ee%
such that (\ref{hol}) becomes $h_{v,I}=c_{v,I}+2\tau_I s_{v,I}$.
In a continuum approximation, we have
\be  \label{contapprox}
c_{v,I}=\cos(\gamma k_I(v+I/2)/2) \quad,\quad s_{v,I}=\sin(\gamma
k_I(v+I/2)/2)
\ee%
where $k_I(v)=\ell_0 \t k_I(v)$. After substituting this expression
into (\ref{Kcontrib}) and making use of the identity\footnote{Here the
fundamental representation of $\tau_I$ has been used: $\tr(\id)=2,
\,\tr(\tau_I \tau_J)=-\f{1}{2}\delta_{IJ}$} (for some fixed
$I,J,K$ and numbers $x_i$ and $y_i$)
\bq%
\epsilon_{IJK}\tr\left[(x_1 \id+2y_1\tau_I )(x_2 \id+
2y_2\tau_J)(x_3\id+2y_3\tau_K) \right]&=&\epsilon_{IJK} \tr (x_1 x_2
x_3  \id )+8\epsilon_{IJK} \tr (y_1 y_2 y_3 \tau_I \tau_J \tau_K)
\nonumber
\\&=&2 (x_1 x_2 x_3 - y_1 y_2 y_3 \epsilon_{IJK})\epsilon_{IJK},
\nonumber
\eq%
any one term of the sum in (\ref{Kcontrib}) becomes
\bq\label{one_sum}%
&&\frac{i}{8\pi\gamma G\hbar} \tr(h_{v,I} h_{v+I,J}
h_{v+J,I}^{-1} h_{v,J}^{-1} h_{v,K}[h_{v,K}^{-1},\hat{V}_v])\\
&=&-\epsilon_{IJK} \left\{\left[(c_{v,I}
c_{v+J,I}+s_{v,I} s_{v+J,I})c_{v,J}
c_{v+I,J}+(c_{v,I} c_{v+J,I}-s_{v,I}
s_{v+J,I})s_{v,J}
s_{v+I,J}\right]\hat A_{v,K}\right\} \nonumber \\
&&+\epsilon_{IJK}^2 \left\{\left[(c_{v,I}
s_{v+J,I}-s_{v,I} c_{v+J,I})s_{v,J} c_{v+I,J}+(s_{v,I}
c_{v+J,I}+c_{v,I} s_{v+J,I})c_{v,J} s_{v+I,J}\right]\hat
B_{v,K}\right\},\nonumber
\eq%
where
\begin{eqnarray}\label{AB_def}%
\hat A_{v,K} &:=& \f{1}{4\pi i \gamma G\hbar} \left(\hat V_v -
c_{v,K} \hat V_v c_{v,K} - s_{v,K}\hat V_v s_{v,K}\right)\,,\nonumber\\
\hat B_{v,K} &:=& \f{1}{4\pi i \gamma G\hbar} \left(s_{v,K} \hat V_v
c_{v,K} - c_{v,K}\hat V_v s_{v,K}\right)\,.
\end{eqnarray}
In the first line of (\ref{one_sum}), the expression inside the
curly braces is symmetric in the indices $I$ and $J$, hence
vanishes when contracted with $\epsilon_{IJK}$. Therefore only the
second line contributes, and the extrinsic curvature part of the
gravitational constraint is
\bq\label{K_operator}%
\hat{H}_{K,v} &=& \frac{-N(v)}{64\pi\gamma^2 G}
 \sum_{IJK}\sum_{\sigma_I\in\{\pm1\}} \{[(c_{v,\sigma_II}
s_{v+\sigma_JJ,\sigma_II}-s_{v,\sigma_II}
c_{v+\sigma_JJ,\sigma_II})s_{v,\sigma_JJ}
c_{v+\sigma_II,\sigma_JJ}\nonumber\\
&&\qquad\qquad+(s_{v,\sigma_II}
c_{v+\sigma_JJ,\sigma_II}+c_{v,\sigma_II}
s_{v+\sigma_JJ,\sigma_II})c_{v,\sigma_JJ} s_{v+\sigma_II,\sigma_JJ}
]\hat B_{v,\sigma_KK}\} \\
&=& \frac{-N(v)}{64\pi\gamma^2 G}
\sum_{IJK}\sum_{\sigma_I\in\{\pm1\}}
\{[ s_{v,\sigma_II,\sigma_JJ}^-
s_{v,\sigma_JJ} c_{v+\sigma_II,\sigma_JJ}
+ s_{v,\sigma_II,\sigma_JJ}^+
c_{v,\sigma_JJ}
s_{v+\sigma_II,\sigma_JJ}]\hat B_{v,\sigma_KK}\}\,, \nonumber
\eq%
where in the last line trigonometric identities have been used
to express products of sines and cosines through
\[
 s_{v,\sigma_II,\sigma_JJ}^{\pm}:=
\sin\left(\frac{\gamma}{2}(k_{\sigma_II}(v+\sigma_II/2)\pm 
k_{\sigma_II}(v+\sigma_JJ+\sigma_II/2)\right)\,.
\]
As in this expression, all functions $k_I$ are, as before, evaluated
at the midpoint of the edge $e_{v,I}$. We see that in the homogeneous
case the first term in the sum vanishes and the leading contribution
is
\be \label{K_hom} 4\sin (\gamma k_I/2) \cos (\gamma k_I/2)
\sin (\gamma k_J/2)
\cos (\gamma k_J/2) \hat B_{v,K},\ee
in agreement with \cite{HomCosmo}.

\subsubsection{Higher curvature corrections}
\label{HigherCurv}

There are two types of corrections visible from this expression: Using
commutators to quantize inverse densitized triad components implies
eigenvalues of $\hat{B}_{v,I}$ which differ from the classical
expectation at small labels $\mu_{v,I}$. Moreover, using holonomies
contributes higher order terms in extrinsic curvature together with
higher spatial derivatives when sines and cosines are expanded in
small curvature regimes. We will now compute the next-leading terms of
higher powers and spatial derivatives of $\tilde{k}_I(v)$ before
dealing with inverse power corrections in the following subsection.

First, recall the usual expectation that quantum gravity gives rise to
low energy effective actions with higher curvature terms such as
$\int\md^4x \sqrt{|\det g|} \ell_{\rm P}^2R^2$ or $\int\md^4x
\sqrt{|\det g|}\ell_{\rm P}^2R_{\mu\nu\rho\sigma}
R^{\mu\nu\rho\sigma}$ added to the Einstein--Hilbert action
$\int\md^4x\sqrt{|\det g|}R$. Irrespective of details of numerical
coefficients, there are two key aspects: The Planck length $\ell_{\rm
P}=\sqrt{G\hbar}$ must be involved for dimensional reasons in the
absence of any other length scale, and higher spatial as well as time
derivatives arise with higher powers of $R_{\mu\nu\rho\sigma}$. In
canonical variables, one expects higher powers and higher spatial
derivatives of extrinsic curvature and the triad, together with
components of the inverse metric necessary to define scalar quantities
from higher curvature powers (which forces one to raise indices on the
Riemann tensor, for instance). Higher time derivatives, on the other
hand, are more difficult to see in a canonical treatment and
correspond to the presence of independent quantum variables without
classical analog
\cite{Karpacz}.

Any quantization such as that followed here starts from the purely
classical action where $\hbar$ and thus $\ell_{\rm P}$ vanishes.  In
effective equations of the resulting quantum theory, quantum
corrections depending on $\hbar$ will nevertheless emerge. As a first
step in deriving such effective equations, we have non-local holonomy
terms in a Hamiltonian operator which through its expectation values
in semiclassical states will give rise to similar contributions of the
same functional form of $k_I(v)$. At first sight, however, the
expressions above do not agree with expectations from higher curvature
actions: One can easily see that in (\ref{K_operator}) there are
higher powers of extrinsic curvature by expanding the trigonometric
functions, and higher spatial derivatives of extrinsic curvature by
Taylor expanding the discrete displacement involved, e.g., in
$k_I(v+I/2)$.  Moreover, higher spatial derivatives of the triad arise
from similar non-local terms in the spin connection contribution
$\hat{H}_{\Gamma}$ discussed later.  But there are no factors of the
Planck length in such higher powers (all factors of $G$ and $\hbar$
are written out explicitly and not ``set equal to one''). In fact, by
definition $k_I(v)$ is dimensionless since it is obtained by
multiplying the curvature component $\tilde{k}_I(v)$ with $\ell_0$ in
which all possible dimensions cancel. Higher power terms here thus do
not need any dimensionful prefactor. Moreover, there are no components
of the inverse metric (which would be $1/\tilde{p}^{I}(v)$ for our
diagonal triads) in contrast to what is required in higher curvature
terms.

\paragraph{Curvature expansion.}

To see how this is reconciled, we expand the Hamiltonian explicitly in
$\ell_0$ after writing $k_I=\ell_0\tilde k_I$. This corresponds to a
slowly varying field approximation with respect to the lattice
size. For the $(+I,+J)$-plaquette, a single term in the sum
(\ref{Kcontrib}) becomes
\begin{eqnarray} \label{K_expansion}
 &&
2s_{v,I,J}^-
s_{v,J}
c_{v+I,J}+
2s_{v,I,J}^+
c_{v,J}
s_{v+I,J}
= \gamma^2\ell_0^2 \tilde{k}_I\tilde{k}_J+
\f{1}{2}\gamma^2 \ell_0^3\left(\tilde{k}_I \tilde{k}_{J,J}+\tilde{k}_J  \tilde{k}_{I,I}+2\tilde{k}_J \tilde{k}_{I,I}\right)\nonumber\\
&& \qquad\qquad+\frac{1}{8}\gamma^2\ell_0^4\left(\t k_I \t k_{J,JJ} +\t k_J \t
k_{I,II}+4(\t k_I\t k_{J,II}+\t k_I\t k_{J,IJ}+\t k_{I,I}\t
k_{J,I}+\t k_{I,J}\t k_{J,I}) \right.\nonumber \\ 
&& \qquad\qquad\left. +2\t k_{I,I}\t
k_{J,J}- \f{4}{3}\gamma^2\t k_I\t k_J(\t k_I^2+\t k_J^2)\right)+
O(\ell_0^5)\,.
\end{eqnarray}
(Commas on the classical field $\tilde{k}_I$ indicate partial
derivatives along a direction given by the following index.)
For a fixed direction $K$ there are in total eight terms to
be included in the sum (\ref{Kcontrib}). They are obtained from
(\ref{K_expansion}) by taking into account the four plaquettes in
the $(I,J)$-plane meeting at vertex $v$ (Fig. \ref{4plaquette})
and considering both orientations in which each plaquette can be
traversed. While the latter merely boils down to symmetrization
over $I$ and $J$, the former requires some care, noting that in the
Hamiltonian constraint (\ref{Kcontrib}) $h_{v,-I}$ means
$h_{v-I,I}^{-1}$. The contribution (\ref{K_expansion}) corresponds to
plaquette 1 of Fig. \ref{4plaquette} and has
$\sigma_I=\sigma_J=\sigma_K=1$. Accounting for the overall sign
dictated by the $\sigma$-factors, one can obtain the expressions for
the three remaining plaquettes 2, 3 and 4 following the recipe below.
\[
\begin{array}{c|cccc|l}
\mbox{plaquette} & \multicolumn{4}{c|}{\mbox{extrinsic curvature
components}} & \mbox{sign}\\\hline
(1)&  k_I(v+I/2)&  k_J(v+I+J/2)&  -k_I(v+I/2+J)& -k_J(v+J/2)&     \\
(2)& -k_I(v-I/2)&  k_J(v-I+J/2)&   k_I(v-I/2+J)& -k_J(v+J/2)&(-1)\\
(3)& -k_I(v-I/2)& -k_J(v-I-J/2)&   k_I(v-I/2-J)&  k_J(v-J/2)&(-1)^2 \\
(4)&  k_I(v+I/2)& -k_J(v+I-J/2)&  -k_I(v+I/2-J)& k_J(v-J/2)&(-1)
\end{array}
\]
The first column designates a plaquette number, whereas the last
one indicates the overall sign factor. The other four columns show
the correspondence between the relevant link labels.

\setlength{\unitlength}{0.01\linewidth}%
\setlength{\fboxsep}{0pt}%
\setlength{\fboxrule}{1.5pt}
\begin{figure}
\begin{center}
\begin{picture}(20,20)
        \linethickness{1pt}
        \put(\LL,\LL){\circle*{2}}
            \put(0,0){\circle*{1}}
            \put(0,\LL){\circle*{1}}
            \put(\LL,0){\circle*{1}}
            \put(\LT,0){\circle*{1}}
            \put(0,\LT){\circle*{1}}
            \put(\LL,\LT){\circle*{1}}
            \put(\LT,\LL){\circle*{1}}
            \put(\LT,\LT){\circle*{1}}
        \put(0,0){\line(1,0){\LT}}
            \put(0,0){\line(0,1){\LT}}
            \put(0,\LL){\line(1,0){\LT}}
            \put(\LL,0){\line(0,1){\LT}}
            \put(\LT,0){\line(0,1){\LT}}
            \put(0,\LT){\line(1,0){\LT}}
        \put(+11,8){$v$}
            \put(-6,10){$v\!-\!I$}
            \put(+21,10){$v\!+\!I$}
            \put(7,-3){$v-J$}
            \put(7,21){$v+J$}
            \put(+21,20){$v\!+\!\!I\!\!+\!\!J$}
            \put(+21,0){$v\!+\!\!I\!\!-\!\!J$}
            \put(-8,20){$v\!\!-\!\!I\!\!+\!\!J$}
            \put(-8,0){$v\!\!-\!\!I\!\!-\!\!J$}
        \put(14.5,14){\large{\bf 1}}
            \put(4.5,14){\large{\bf 2}}
            \put(4.5,4){\large{\bf 3}}
            \put(14.5,4){\large{\bf 4}}
        \put(\LL,\LL){\vector(1,0){6}}
            \put(\LL,\LL){\vector(-1,0){6}}
            \put(\LT,\LL){\vector(0,1){6}}
            \put(\LT,\LL){\vector(0,-1){6}}
            \put(0,\LL){\vector(0,1){6}}
            \put(0,\LL){\vector(0,-1){6}}
            \put(0,0){\vector(1,0){6}}
            \put(0,\LT){\vector(1,0){6}}
            \put(\LL,0){\vector(0,1){6}}
            \put(\LL,\LT){\vector(0,-1){6}}
            \put(\LT,0){\vector(-1,0){6}}
            \put(\LT,\LT){\vector(-1,0){6}}

\end{picture}
\end{center}
 \caption{Four plaquettes adjacent to vertex $v$ in the $(I,J)$-plane.
 The arrows indicate the directions in which the relevant holonomies are traversed. \label{4plaquette}}
\end{figure}
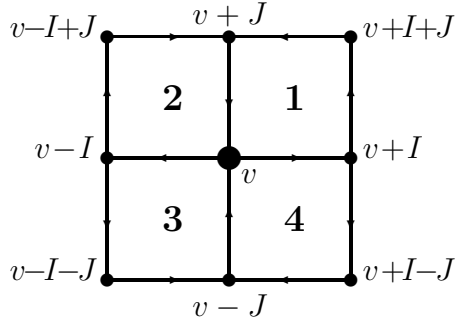
After the symmetrization over all four plaquettes (traversed in
both directions), the cubic terms drop out
\begin{eqnarray} \label{K_ave}
\gamma^2\ell_0^2 \tilde{k}_I\tilde{k}_J &\!\!-\!\!&\f{\gamma^4
\ell_0^4}{6} \t k_I \t k_J (\t k_I^2+\t k_J^2)
\\&\!\!+\!\!&\frac{\gamma^2 \ell_0^4}{8}\left(\t k_I \t k_{J,JJ} +\t k_J
\t k_{I,II}+2(\t k_I\t k_{J,II}+\t k_J\t k_{I,JJ}+\t k_{I,I}\t
k_{J,I}+\t k_{I,J}\t k_{J,J}) \right) + O(\ell_0^5)\,.\nonumber
\end{eqnarray}
Note that the link labels $\t k$ were introduced as values of the
extrinsic curvature components evaluated at midpoints of edges in the
continuum approximation (\ref{contapprox}) of our basic non-local
variables. The expression above is written in terms of just two
components $\t k_I(v)$ and $\t k_J(v)$ (and their partial
spatial derivatives) Taylor expanded around the vertex $v$.  The first
term, when combined with $\hat{B}_{v,K}$ and summed over all triples $IJK$,
reproduces the correct classical limit of the constraint $H_K$. This
limit is obtained in two steps: we first performed the continuum
approximation by replacing holonomies with mid-point evaluations of
extrinsic curvature components. This would still give us a non-local
Hamiltonian since each vertex contribution now refers to evaluations
of the classical field at different points. In a second step we then
Taylor expanded these evaluations around the central vertex $v$, which
gives a local result and corresponds to a further, slowly-varying
field approximation.

\paragraph{Comparison with higher curvature terms.}

Here, the factor $\ell_0^2$ in the leading term together with a factor
$\ell_0$ from $\hat{B}_{v,K}$ through (\ref{commPoiss}) combines to
give the Riemann measure of the classical integral. Higher order
terms, however, come with additional factors of $\ell_0$ in
(\ref{K_ave}) which are not absorbed in this way.  The result is
certainly independent of coordinates since the whole construction
(\ref{K_operator}) in terms of $k_I$ is coordinate independent. But
for a comparison with higher curvature terms we have to formulate
corrections in terms of $\tilde{k}_I$ and $\tilde{p}^I$ as these are
the components of classical extrinsic curvature and densitized triad
tensors. Higher order terms in the expansion are already formulated
with $\tilde{k}_I$ in coordinate independent combinations with
$\ell_0$-factors. It remains to interpret the additional $\ell_0$
factors appropriately for a comparison with low energy effective
actions.

This can be done quite simply in a way which removes the above
potential discrepancies between our expansions and higher curvature
terms in low energy effective actions. We simply use
(\ref{ScalarFlux}) to write $\ell_0^2=p^I/\tilde{p}^I$ which is the
only well-defined possibility to express $\ell_0$ in terms of the
fields. Thus, inverse metric components $1/\tilde{p}^I$ directly occur
in combination with $\tilde{k}_J$ factors as required for higher
curvature terms. The fact that the cubic term in $\ell_0$ in
(\ref{K_ave}) drops out is also in agreement with higher curvature
corrections since in that case only even powers of the length scale
$\ell_{\rm P}$ occur. Moreover, there are now factors of $p^I$
multiplying the corrections. These are basic variables of the quantum
theory determining the fundamental discreteness. Thus, factors of the
Planck length occurring in low energy effective actions are replaced
by the state specific quantities $p^I$. While the Planck length
$\ell_{\rm P}=\sqrt{G\hbar}$ is expected to appear for dimensional
reasons without bringing in information about quantum gravity (it can
just be computed using classical gravity for $G$ and quantum mechanics
for $\hbar$), the $p^I$ are determined by a state of quantum
gravity. If expressed through labels $\mu_{v,I}$, the Planck length
also appears, but it can be enlarged when $\mu_{v,I}>1$. Moreover, the
lattice labels are dynamical (and in general inhomogeneous) and can
thus change in time in contrast to $\ell_{\rm P}$. Although the form
of corrections is analogous to those for low energy effective actions,
the conceptual as well as dynamical appearance of correction terms is
thus quite different.

The terms considered so far could not give rise to higher time
derivatives of the spatial metric. In general, higher time derivatives
describe the effect of quantum variables (\ref{QuantVars}) of the
field theory, which appear in the expectation value of the Hamiltonian
constraint in a generic state. Quantum variables are thus, in a
certain sense, analogous to higher time derivatives in effective
actions \cite{Karpacz}, which indicates that the correction terms they
imply should combine with those in (\ref{K_ave}) obtained by
expanding sines and cosines to higher powers of space-time curvature
components. All corrections of these types should thus be considered
together since they will eventually be mixed up despite of their
different derivations. A computation of terms containing quantum
variables requires more detailed information about the expectation
value of the constraint operator in an arbitrary state. These terms
are thus more difficult to compute, which also makes an interpretation
of the remaining higher curvature terms alone, especially concerning
their possible covariance, more difficult.\footnote{It is sometimes
tempting to ``sum the whole perturbations series'' of higher order
terms simply by using the left hand side of (\ref{K_expansion})
directly in effective equations without an expansion. However, this is
in general not a consistent approximation since arbitrarily small
higher order terms are included but other types of corrections such as
higher time derivatives or quantum variables are completely
ignored. There is currently only one known model where the procedure
is correct since all quantum variables have been shown to decouple
from the expectation values \cite{BouncePert}. But this model, a free
isotropic scalar in a certain factor ordering of the constraint
operator, is very special. Under modifications such as a scalar
potential quantum variables do not decouple and their motion implies
further correction terms in effective equations not captured in the
trigonometric functions arising from holonomies.}  We will thus focus
from now on on corrections coming from commutators $\hat{B}_{v,K}$ to
quantize inverse powers which are independent of the higher order
corrections and even give rise to non-perturbative terms. Moreover, in
Sec.~\ref{magnitude} below we will demonstrate that those corrections
are expected to be dominant in cosmological perturbation theory.

\subsubsection{Inverse triad corrections}

A direct calculation using (\ref{hol_action}) and (\ref{V_action})
shows that $\hat B_{v,K}$ commutes with all flux operators and thus
has flux eigenstates as eigenbasis, as it happens also in
homogeneous models \cite{LivRev}. The action
\begin{eqnarray}
\hat B_{v,K}|\ldots,\mu_{\vertex,K},\ldots\rangle &:=&
\left(2\pi\gamma
\lP^2\right)^{1/2}\sqrt{|\mu_{\vertex,I}+\mu_{\vertex,-I}|
|\mu_{\vertex,I}+\mu_{\vertex,-J}|} \\
&\times&\left(\sqrt{|\mu_{\vertex,K}+\mu_{\vertex,-K}+1|}-
\sqrt{|\mu_{\vertex,K}+\mu_{\vertex,-K}-1|}\right)
|\ldots,\mu_{\vertex,K},\ldots\rangle
\nonumber
\end{eqnarray}
directly shows the eigenvalues which do not agree exactly with the
classical expectation $e_{K}(v)=\sqrt{|p^I(v)p^J(v)/p^K(v)|}\sim
\sqrt{|\mu_{v,I}\mu_{v,J}/\mu_{v,K}|}$ (indices such that $\epsilon^{IJK}=1$)
for the co-triad (\ref{cotriad}) which appears as a factor in the
Hamiltonian constraint. But for large values $\mu_{v,I}\gg1$ the
classical expectation is approached as an expansion of the eigenvalues
shows.

Inverse triad corrections are obtained by extracting the corrections
which $B_{v,K}$ receives on smaller scales.  We introduce the
correction function as a factor $\alpha_{v,K}$, depending on the
lattice labels $\mu_{v,I}$, such that $B_{v,K}=\alpha_{v,K}
e_{v,K}$ and $\alpha_{v,K}\rightarrow 1$ classically, i.e.\ for
$\mu_{v,K}\gg1$.
Comparing the eigenvalues of $\hat B_{v,K}$ with
those of flux operators in the combination $\sqrt{|{\cal F}_{v,I}{\cal
F}_{v,J}/{\cal F}_{v,K}|}$, we find
\be \label{alpha_corr}
\alpha_{v,K}=\sqrt{|\mu_{\vertex,K}+\mu_{\vertex,-K}|}
\left(\sqrt{|\mu_{\vertex,K}+\mu_{\vertex,-K}+1|}-
\sqrt{|\mu_{\vertex,K}+\mu_{\vertex,-K}-1|}\right)\ee

After having computed the operators and their eigenvalues, we can
specialize the correction function to perturbations of the scalar
mode. We reduce the number of independent labels by imposing
$\mu_{v,I}+\mu_{v,-I}=\mu_{v,J}+\mu_{v,-J}$ for arbitrary $I$ and
$J$. This corresponds to a metric proportional to the identity
$\delta_{ab}$ for a scalar perturbation. We then assign a new variable
$p(v)=2\pi\gamma\ell_{\rm P}^2(\mu_{v,I}+\mu_{v,-I})$ to each vertex
$v$, which is independent of the direction of the edge $I$ and
describes the diagonal part of the triad. Quantum numbers in
eigenvalues of the lattice operators can then be replaced by $p(v)$,
and the resulting functions compared with the classical ones. The
remaining subscript $v$ indicates that the physical quantities are
vertex-dependent, i.e.\ inhomogeneous. Then the averaging over the
plaquette orientations in the constraint becomes trivial and the
total correction reads
\be\label{alpha_iso}
\alpha_v=2\sqrt{|\mu_v|}\left(\sqrt{|\mu_v+1/2|}-\sqrt{|\mu_v-1/2|}\right)
\ee
i.e.\
\begin{equation}
 \alpha[p(v)]=\frac{\sqrt{|p(v)|}}{2\pi\gamma\ell_{\rm P}^2}
\left(\sqrt{|p(v)+2\pi\gamma\ell_{\rm P}^2|}-
\sqrt{|p(v)-2\pi\gamma\ell_{\rm P}^2|}\right)\,.
\end{equation}
We will continue analyzing these correction functions in
Sec.~\ref{CorrProp} after having discussed how such functions also
enter the spin connection and matter terms.

\subsubsection{Spin connection}
\label{SpinConn}

So far, the holonomies we used only contributed the extrinsic
curvature terms to the Hamiltonian but no spin connection terms at
all. In the procedure followed here, we thus have to quantize
$\Gamma_a^i[E]$ directly which is possible in the perturbative regime
where line integrals of the spin connection have covariant meaning.
This gives rise to one further correction function in the effective
expression of the spin connection
\be%
\Gamma_I^i=-\epsilon^{ijk} e_j^b \left(\partial_{[I}
e_{b]}^k+\f{1}{2}e_k^c e_a^l \partial_{[c} e_{b]}^l\right), \ee
as it also contains a co-triad (\ref{cotriad}). Since the triad and its
inverse have a diagonal form
\be%
e_i^I\equiv\f{E_i^I}{\sqrt{|\det E|}}=e^{(I)}\delta_i^I, \quad
e_I=e_{(I)}\delta_I^i
\ee%
with the components given by%
\be \label{e_comp} e^{I}=\f{p^I}{\sqrt{|\det E|}}=\left(
e_{I}\right)^{-1} ,\quad \det E = p^I p^J
p^K, \ee%
the spin-connection simplifies to
\be \label{Spinscalar}
\Gamma_I^i=\epsilon^{ic}_I e^{(c)}\partial_c e_{(I)}.\ee
In terms of components of a densitized triad it reads
\be\label{spin_class}%
\Gamma_I^i=\f{1}{2}\epsilon^{ij}_I\f{p^{(j)}}{p^{(I)}}
\left(\sum_J\f{\partial_j p^J}{p^J}-2 \f{\partial_j p^I}{p^I}\right).%
\ee%

Since there are many alternative choices in performing the
quantization of such an object, but not much guidance from a potential
operator in the full theory, we first discuss general aspects one can
expect for the quantization of the spin connection in a simple
version. It includes corrections of inverse densitized triad
components by correction functions in each term of (\ref{spin_class}).
We thus mimic a quantization to the extent that expectation values of
classical expressions containing inverse powers of $p$ acquire a
correction factor
\be
\f{1}{p^I(v)} \rightarrow \f{\beta_I(v)}{p^I(v)},\ee where the
correction functions $\beta_I$ are kept different from the function
$\alpha$ used before because the object to be quantized is different.
There will also be corrections from the discretization $\Delta_I$ of
partial derivatives $\ell_0\partial_I$, but we ignore them in what
follows for the same reason which allowed us to ignore such effects
from the loop holonomy quantizing $F_{ab}^i$.  The effective analog
of (\ref{spin_class}) is then of the form
\be\label{spin_eff}%
(\Gamma_I^i)_{\rm
eff}=\f{1}{2}\epsilon^{ij}_I\beta^I\f{p^j}{p^I}\left(\sum_J\beta^J\f{\partial_j
p^J}{p^J}-2\beta^I \f{\partial_j p^I}{p^I}\right).%
\ee%
At this stage the triad components, corresponding to different
orientations, can be put equal to each other in effective equations,
$p^I=p^J=p^K=p$. This implies an analogous relation between the
correction functions $\beta_I=\beta_J=\beta_K=\beta_0$. Comparing
(\ref{spin_eff}) with the ansatz $\Gamma_I^i=\f{1}{2}\epsilon^{ij}_I
\beta \f{\partial_j p}{p}$, we conclude that also the spin connection
receives a correction function $\beta=\beta_0^2$.

For a precise quantization we observe that we need terms of the form
$\ell_0^2\Gamma_a^i\Gamma_b^j$ and $\ell_0^2\partial_a\Gamma_b^i$ in
the constraint since one factor $\ell_0$ of the Riemann measure will
be absorbed in the commutator $\hat{B}_{v,I}$. To quantize
$\ell_0\Gamma_a^i$, we combine $\ell_0$ with the partial derivative
$\partial_I$ in (\ref{Spinscalar}) to approximate a lattice difference
operator $\Delta_I$ defined by $(\Delta_If)_v= f_{v+I}-f_v$ for any
lattice function $f$. A well-defined lattice operator thus results
once a prescription for quantizing the inverse triad has been
chosen. One can again make use of Poisson identities for the classical
inverse which, however, allows more freedom than for the combination
of triad components we saw in the Hamiltonian constraint. Such a
freedom, corresponding to quantization ambiguities, will also be
encountered when we consider matter Hamiltonians. For any choice we
obtain a well-defined operator which would not be available without
the perturbative treatment since the full spin connection is not a
tensorial object.

An explicit example can most easily be derived by writing the spin
connection integrated along a link $e_{v,I}$ as it might appear in a holonomy,
\[
 \int_{e_{v,I}} \dot{e}_I^a\Gamma^i_a\approx \ell_0\Gamma^i_I
 =\epsilon^{ic}_I e^{(c)}\ell_0
\partial_c e_{(I)}\approx
\epsilon^{iK}_I \frac{p^{(K)}}{\sqrt{|\det E|}} \Delta_K e_{(I)}
\]
using the lattice difference operator
$\Delta_I\approx\ell_0\partial_I$. We then have to deal with the
inverse powers explicit in the fraction and implicit in the co-triad
$e_I$. The latter is standard, replacing $e_I$ by
$\ell_0^{-1}h_I\{h_I^{-1},V_v\}$ based on (\ref{cotriad}). The inverse
determinant in the fraction cannot be absorbed in the resulting
Poisson bracket because (i) it does not commute with the derivative
and (ii) absorbing a single inverse in a single co-triad would lead to
a logarithm of $V_v$ in the Poisson bracket which would not be
well-defined. It can, however, be absorbed in the flux $\ell_0^2 p^K$
if we do not use the basic flux operator $\hat{F}_{v,K}$ but the
classically equivalent expression
\begin{eqnarray}
 F_{v,K} &\approx& \ell_0^2 p^K= \frac{1}{2}\ell_0^2 \delta^k_{(K)}
\epsilon_{kij}\epsilon^{KIJ} e_I^ie_J^j\nonumber\\
&=&-\frac{1}{4}(4\pi\gamma G)^{-2}\sum_{IJ}\sum_{\sigma_I\in\{\pm1\}}
\sigma_I\sigma_J
\epsilon^{IJK}\tr(\tau_{(K)} h_I\{h_I^{-1},V_v\} h_J\{h_J^{-1},V_v\})
\end{eqnarray}
which is analogous to expressions used in \cite{Flux}. Since there are
two Poisson brackets, we can split the inverse $V_v$ evenly among
them, giving rise to square roots of $V_v$ in the brackets:
\begin{eqnarray}
 \frac{p^K}{\sqrt{|\det E|}} &\approx&
\ell_0\frac{F_{v,K}}{V_v}\nonumber\\
&=& -\frac{\ell_0}{16\pi^2\gamma^2 G^2}
\sum_{IJ}\sum_{\sigma_I\in\{\pm1\}}\sigma_I\sigma_J
\epsilon^{IJK}\tr(\tau_{(K)} h_I\{h_I^{-1},\sqrt{V_v}\}
h_J\{h_J^{-1},\sqrt{V_v}\})\,.
\end{eqnarray}
The remaining factor of $\ell_0$ is absorbed in $e_I$ inside the
derivative which is quantized following the standard procedure. A
well-defined quantization of spin connection components thus follows,
which is not local in a vertex since the difference operator connects
to the next vertex. Similarly, the derivative of the spin connection
needed in the Hamiltonian constraint leads to further connections to
next-to-next neighbors.

Explicitly, one can thus write an integrated spin connection operator
quantizing $\Gamma_{v,I}^i:=\int_{e_{v,I}}\md t\dot{e}_I^a\Gamma_a^i$ as
\begin{eqnarray}
 \hat{\Gamma}_{v,I}^i &=& \epsilon_I{}^{iK}
\left(\frac{1}{16\pi^2\gamma^2\ell_{\rm P}^2}
\sum_{J,L,\sigma_J,\sigma_L} \sigma_J\sigma_L\epsilon^{JLK}
\tr\left(\tau_{(K)} h_J[h_J^{-1},\hat{V}_v^{1/2}]
h_L[h_L^{-1},\hat{V}_v^{1/2}]\right) \right. \nonumber\\
&&\times\left.\Delta_K\left(\frac{i}{2\pi\gamma\ell_{\rm P}^2}
\tr(\tau^{(I)} h_I[h_I^{-1},\hat{V}_v])\right)\right)\,.
\end{eqnarray}
Replacing the commutators by classical expressions times correction
functions $\alpha$ defined as before and $\alpha_{1/2}$ defined
similarly for a commutator containing the square root of the volume
operator leads to an expression
\begin{eqnarray*}
 (\Gamma_I^i)_{\rm eff} &=&
\alpha_{1/2}(p^i)\alpha_{1/2}(p^I)\epsilon_I{}^{ic}
e^{(c)}\partial_c(\alpha(p^I)e_I) \\
&=&\alpha_{1/2}(p^i)\alpha_{1/2}(p^I)\alpha(p^I) \Gamma_I^i+
\alpha_{1/2}(p^i)\alpha_{1/2}(p^I)
\alpha'(p^I)e_I\epsilon_I{}^{ic}e^{(c)}\partial_cp^{(I)}
\end{eqnarray*}
where the prime denotes a derivative by $p^I$. Using the relation
$p^J=e_Ie_K$ whenever $\epsilon_{JIK}=1$ between densitized triad and
co-triad components allows us to write
\begin{eqnarray}
 (\Gamma_I^i)_{\rm eff} &=&
\alpha_{1/2}(p^i)\alpha_{1/2}(p^I)\alpha(p^I) \Gamma_I^i+
\alpha_{1/2}(p^i)\alpha_{1/2}(p^I)
\alpha'(p^I)e_I\epsilon_I{}^{ic}e^{(c)}\partial_c(e_Je_K)|_{\epsilon_{IJK}=1}
\nonumber\\
&=& \alpha_{1/2}(p^i)\alpha_{1/2}(p^I)
\left(\alpha(p^I)\Gamma^i_I+\sum_{K\not=I}
\alpha'(p^I) p^K\Gamma^i_K\right)
\end{eqnarray}
for the effective spin connection components. For scalar modes, using
that all $p^I$ at a given point are equal, this can be written with a
single correction function
\begin{equation}
 \beta[p(v)]=\alpha_{1/2}[p(v)]^2(\alpha[p(v)]+2p\alpha'[p(v)])
\end{equation}
for $\Gamma_I^i$, where $\alpha'=\md\alpha/\md p$.

\subsection{Matter Hamiltonian}

Matter fields are quantized by similar means in a loop
quantization, using lattice states, and then coupled dynamically
to geometry by adding the matter Hamiltonian to the constraint.
For a scalar field $\phi$, the momentum $\pi=\sqrt{|\det
E|}\dot{\phi}/N$ is a density of weight one. In the
$\phi$-representation, states will simply be of the form already
used for the gravitational field, except that each vertex now
also carries a label $\nu_v\in{\mathbb R}$ describing the
dependence on the scalar field $\phi(v)$ through
$\exp(i\nu_v\phi(v))$ \cite{ScalarBohr}. Well-defined lattice
operators are then given by $\widehat{\exp(i\nu_0\phi}_v)$, for
any $\nu_0\in{\mathbb R}$, which shifts the label $\nu_v$ by
$\nu_0$. The momentum, with its density weight, has to be
integrated before it can meaningfully be quantized. We introduce
\[
 P_v := \int_{R_v}\md^3x \pi\approx \ell_0^3\pi(v)
\]
where $R_v$ is a cubic region around the vertex $v$ of the size of a
single lattice site. Since we have $\{\phi(v),P_w\}= \chi_{R_w}(v)$ in
terms of the characteristic function $\chi_R(v)=1$ if $v\in R$ and
zero otherwise, a momentum operator $P_v$ must have eigenvalue $\hbar\nu_v$
in a state introduced above.

\subsubsection{Inverse triad corrections}

For the matter Hamiltonian of a scalar field $\phi$ with momentum
$\pi$ and potential $U(\phi)$ we have the classical expression
\[
H_{\phi}[N]=\int \md^3x N(x)
\left[\frac{1}{2\sqrt{\det(q)}}\pi(x)^2
+\frac{E_i^aE_i^b}{2\sqrt{\det q}}\partial_a\phi(x)\partial_b
\phi(x)+\sqrt{\det q}U(\phi)\right]
\]
containing inverse powers of the metric, too.
It can be quantized by loop techniques \cite{QSDV,QFTonCSTI} making
use of identities similar to (\ref{cotriad}). One first generalizes
the identity to arbitrary positive powers of the volume in a Poisson
bracket,
\begin{equation}
 \{A_a^i,V_v^r\}=4\pi\gamma G\:rV_v^{r-1}e_a^i
\end{equation}
and then combines such factors with suitable exponents $r$ to produce
a given product of triad and co-triad components. Since such
identities would be used only when inverse components of densitized
triads are involved and a positive power of volume must result in the
Poisson bracket, the allowed range for $r$ is $0<r<2$. Any such
Poisson bracket will be quantized to
\[
 \dot{e}^a_K\{A_a^i,V_v^r\}\mapsto
\frac{-2}{i\hbar \ell_0}\tr(\tau^ih_{v,K}[h_{v,K}^{-1},\hat{V}_v^r])
\]
using holonomies $h_{v,I}$ in direction $I$ with tangent vector
$\dot{e}_K^a$. Since holonomies in our lattice states have
internal directions $\tau_K$ for direction $K$, we can compute the
trace and obtain
\begin{equation}
\widehat{V_v^{r-1}{e}_K^i}= \frac{-2}{8\pi i r\gamma \lP^2 \ell_0}
\sum_{\sigma \in \{\pm 1\}}\sigma\tr(\tau^ih_{v,\sigma K}[h_{v,\sigma
K}^{-1},\hat{V_v}^r]) =\frac{1}{2\ell_0} (\hat{B}_{v,K}^{(r)} -
\hat{B}_{v,-K}^{(r)}) \delta^i_{(K)}
\end{equation}
where, for symmetry, we use both edges touching the vertex $v$
along direction $K$ and $\hat B_{v,K}^{(r)}$ is the generalized
version of (\ref{AB_def}):
\be\label{B_def}%
\hat B_{v,K}^{(r)} := \f{1}{4 \pi i\gamma G \hbar r}\left(s_{v,K}
\hat
V_v^r c_{v,K} - c_{v,K}\hat V_v^r s_{v,K}\right)\ee%

The exponent used for the gravitational part was $r=1$, and $r=1/2$
already occurred in the spin connection, while the scalar Hamiltonians
introduced in \cite{QSDV,QFTonCSTI}, which we closely follow in the
construction of the matter Hamiltonian here, use $r=1/2$ for the
kinetic term and $r=3/4$ for the gradient term. With
\[
 \epsilon^{abc}\epsilon_{ijk} \{A_a^i,V_v^{1/2}\} \{A_b^j,V_v^{1/2}\}
\{A_c^k,V_v^{1/2}\}= (2\pi\gamma G)^3\epsilon^{abc}\epsilon_{ijk}
\frac{e_a^ie_b^je_c^k}{V_v^{3/2}}= \frac{6(2\pi\gamma
G)^3}{\ell_0^3V_v^{1/2}}
\]
for a lattice site volume $V_v\approx \ell_0^3 |\det(e_a^i)|$ and
\[
 \epsilon^{abc}\epsilon_{ijk} \{A_b^j,V_v^{3/4}\} \{A_c^k,V_v^{3/4}\}=
 (3\pi\gamma G)^2\epsilon^{abc}\epsilon_{ijk}
\frac{e_b^je_c^k}{V_v^{1/2}}= 6(3\pi\gamma
G)^2\frac{E^a_i}{V_v^{1/2}}
\]
one can replace the inverse powers in the scalar Hamiltonian as
follows: For the kinetic term, we discretize
\[
 \int\md^3x \frac{\pi^2}{\sqrt{\det q}}\approx \sum_v
\ell_0^3\frac{\pi(v)^2}{\sqrt{\det q(v)}}\approx \sum_v
\frac{{P_v}^2}{V_v}\,.
\]
Then, the classically singular
\begin{equation} \label{invV}
 \frac{1}{V_v}=\left(\frac{\ell_0^3}{6} \epsilon^{abc}\epsilon_{ijk}
\frac{e_a^ie_b^je_c^k}{V_v^{3/2}}\right)^2=
\left(\frac{\ell_0^3}{6(2\pi\gamma G)^3} \epsilon^{abc}\epsilon_{ijk}
\{A_a^i,V_v^{1/2}\} \{A_b^j,V_v^{1/2}\} \{A_c^k,V_v^{1/2}\}\right)^2
\end{equation}
will be quantized to
\[
 \left(\frac{1}{48} \epsilon^{IJK}\epsilon_{ijk}
 (\hat{B}_{v,I}^{(1/2)}- \hat{B}_{v,-I}^{(1/2)})\delta^i_{(I)}
 (\hat{B}_{v,J}^{(1/2)}- \hat{B}_{v,-J}^{(1/2)})\delta^j_{(J)}
 (\hat{B}_{v,K}^{(1/2)}- \hat{B}_{v,-K}^{(1/2)})\delta^k_{(K)}
\right)^2\,.
\]
Similarly, we discretize the gradient term by
\[
 \int\md^3x \frac{E^a_iE^b_i}{\sqrt{\det
q}}\partial_a\phi\partial_b\phi\approx
\sum_v\ell_0^3\frac{E^a_i(v)E^b_i(v)}{\sqrt{\det
q(v)}}(\partial_a\phi)(v)(\partial_b\phi)(v)\approx \sum_v
\frac{p^I(v)p^J(v)}{V_v} \Delta_I\phi_v\Delta_J\phi_v
\]
where we replace spatial derivatives $\partial_a$ by lattice
differences $\Delta_I$. Now, using
\[
 \delta^i_{(I)}\frac{p^I(v)}{V_v^{1/2}}=\ell_0^2\frac{E_i^I(v)}{V_v^{1/2}}=
\frac{\ell_0^2}{6} \frac{\epsilon^{Ibc}\epsilon_{ijk}
e^j_be^k_c}{V_v^{1/2}}= \frac{\ell_0^2}{6(3\pi\gamma G)^2}
\epsilon^{Ibc}\epsilon_{ijk} \{A_b^j,V_v^{3/4}\}\{A_c^k,V_v^{3/4}\}
\]
we can quantize the metric contributions to the gradient term by
\bq
 \frac{1}{24^2}
&\!\!\!\!\!\!\!\!\!\!\!\!\!\!\!\epsilon^{IKL}\epsilon_{ijk}
(\hat{B}_{v,K}^{(3/4)}- \hat{B}_{v,-K}^{(3/4)})\delta^j_{(K)}
(\hat{B}_{v,L}^{(3/4)}- \hat{B}_{v,-L}^{(3/4)})\delta^k_{(L)}
 \label{gradient}\\
&\times\epsilon^{JMN}\epsilon_{imn}
(\hat{B}_{v,M}^{(3/4)}- \hat{B}_{v,-M}^{(3/4)})\delta^m_{(M)}
(\hat{B}_{v,N}^{(3/4)}- \hat{B}_{v,-N}^{(3/4)})\delta^n_{(N)}
\,.\nonumber
 \eq

In addition to the fact that we are using different values for $r$ in
each term in the gravitational and matter parts, giving rise to
different correction functions, the matter terms are less unique than
the gravitational term and can be written with different parameters
$r$. This corresponds to quantization ambiguities which will appear
also in effective equations and which could have phenomenological
implications. Some choices are preferred since they give rise to
simpler expressions, but this does not suffice to determine a unique
quantization. Instead of using $r=1/2$ in the kinetic term, for
instance, we can use the class of relations
\begin{eqnarray*}
  \frac{1}{\sqrt{|\det E|}}=\frac{(\det e)^k}{(\det E)^{(k+1)/2}}=
\left(\frac{1}{6}\epsilon^{abc}\epsilon_{ijk} (4\pi
G\gamma)^3\right.\\
\times
\{A_a^i,V^{(2k-1)/3k}\} \{A_b^j,V^{(2k-1)/3k}\}
\{A_c^j,V^{(2k-1)/3k}\}\Bigr)^k
\end{eqnarray*}
for any positive integer $k$ to write the inverse determinant through
Poisson brackets not involving the inverse volume (see also the
appendix of \cite{Robust}). This determines an integer family of
quantizations with $r_k=(2k-1)/3k>\frac{1}{3}$. For $k=2$ we obtain
the previous expression, but other choices are possible. Moreover,
using the same $r$ in all terms arising in gravitational and matter
Hamiltonians can only be done in highly contrived ways, if at
all. There is thus no clearly distinguished value. From now on we will
work with the choices specified above.

On regular lattice states, all ingredients are composed to a
Hamiltonian operator
\begin{eqnarray} \label{MatterHam}
\hat{H}_{\phi}[N]&=&\sum_{v\in\gamma}
N_v\left[\frac{1}{2}\hat{P}_v^2\left(\f{1}{48}\sum_{IJK,\sigma_I\in\{\pm1\}}
\sigma_1\sigma_2\sigma_3\epsilon^{IJK}
\hat{B}_{v,\sigma_1I}^{(1/2)} \hat{B}_{v,\sigma_2J}^{(1/2)}
\hat{B}_{v,\sigma_3K}^{(1/2)}
\right)^2\right.\\
&+&\left. \frac{1}{2} \left(\f{1}{48}\sum_{IJK,\sigma_I\in\{\pm1\}}
\sigma_1\sigma_2\sigma_3\epsilon^{IJK}
(\sigma_1\Delta_{\sigma_1I}\phi)_v \hat{B}_{v,\sigma_2J}^{(3/4)}
\hat{B}_{v,\sigma_3K}^{(3/4)} \right)^2 +\hat{V}_v U(\phi_v)
\right]\,, \nonumber
\end{eqnarray}

\subsubsection{Matter correction functions}

\begin{figure}
\centerline{\includegraphics[width=10cm,keepaspectratio]{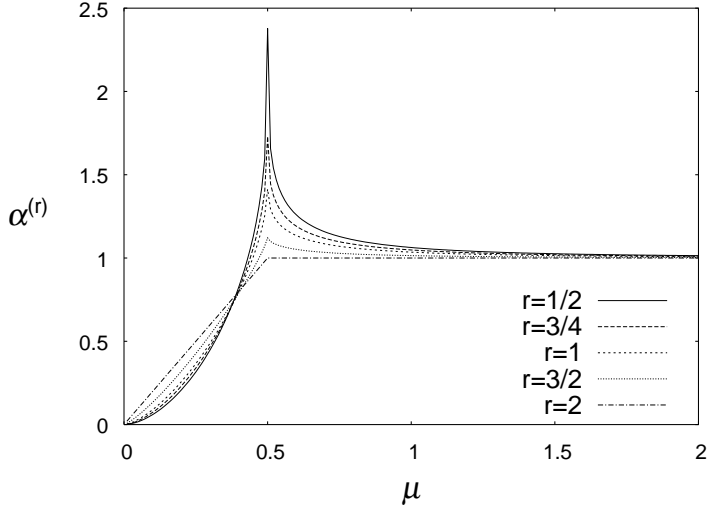}}
\caption{Behavior of the correction function $\alpha$. It
approaches one from above for large arguments. For small arguments,
the function is increasing from zero and reaches a peak value larger
than one. Also shown is the limiting case $r=2$ which does not show a
peak but a constant correction function for
$\mu>1$. \label{Fig_alpha}}
\end{figure}

As before, we compute eigenvalues of the operators \be \hat
B_{v,K}^{(r)}:=\left(2\pi\gamma \lP^2\right)^{-1}\f{\hat
{V}^r|_{\mu_{v,K}+1}-\hat {V}^r|_{\mu_{v,K}-1}}{r}
\ee%
where the subscript of the volume operator indicates that its eigenvalue
in a lattice state is computed according to (\ref{V_action}) with
a shifted label of link $e_{v,I}$. Thus, the eigenvalues are
\be\label{Br_def}%
B_{v,K}^{(r)}:=\frac{1}{r}\left(2\pi\gamma
\lP^2\right)^{\f{3r}{2}-1} |\mu_{v,I}+\mu_{v,-I}|^{r/2}
|\mu_{v,J}+\mu_{v,-J}|^{r/2}
\left(|\mu_{v,K}+\mu_{v,-K}+1|^{r/2}-
|\mu_{v,K}+\mu_{v,-K}-1|^{r/2}\right)\,.
\ee%
compared to the classical expectation
\[
 (2\pi\gamma\ell_{\rm P}^2)^{\f{3r}{2}-1}
|\mu_{v,I}+\mu_{v,-I}|^{r/2} |\mu_{v,J}+\mu_{v,-J}|^{r/2}
|\mu_{v,K}+\mu_{v,-K}|^{r/2-1}
\]
for $V^{r-1}e_K$. For any $r$, correction functions
\begin{eqnarray}
 \alpha_{v,K}^{(r)} &=& \frac{1}{r} |\mu_{v,K}+\mu_{v,-K}|^{1-r/2}
\left(|\mu_{v,K}+\mu_{v,-K}+1|^{r/2}-
|\mu_{v,K}+\mu_{v,-K}-1|^{r/2}\right)
\end{eqnarray}
result. The main examples of $r$ are seen in Fig.~\ref{Fig_alpha}.

Imposing again that $\mu_{v,I}+\mu_{v,-I}=\mu_{v,J}+\mu_{v,-J}$ for
arbitrary $I$ and $J$ and introducing $p(v)=2\pi\gamma\ell_{\rm
P}^2 (\mu_{v,I}+\mu_{v,-I})$, we obtain the effective correction
functions
\begin{eqnarray}
 \alpha^{(r)}[p(v)] &=& \frac{2}{2\pi r\gamma\ell_{\rm P}^2}
|p(v)|^{1-r/2}
\left(|p(v)+2\pi\gamma\ell_{\rm P}^2|^{r/2}-
|p(v)-2\pi\gamma\ell_{\rm P}^2|^{r/2}\right)\,.\nonumber
\end{eqnarray}
This can be used to write the effective matter Hamiltonian on a
conformally flat space $q_{ab}=|p(x)|\delta_{ab}$ as
\[
 H_{\phi}=\int_{\Sigma} \md^3x
N(x)\left(\frac{D[p(x)]}{2|\tilde{p}(x)|^{3/2}} \pi(x)^2
+\frac{\sigma[p(x)])|\tilde{p}(x)|^{\frac{1}{2}}\delta^{ab}}{2}
\partial_a\phi\partial_b\phi
+|\tilde{p}(x)|^{\frac{3}{2}}U(\phi)\right),
\]
where comparison with (\ref{MatterHam}) shows that
\begin{equation}
 D[p(v)]=\alpha^{(1/2)}[p(v)]^6\quad\mbox{ and
}\quad\sigma[p(v)]=\alpha^{(3/4)}[p(v)]^4\,.
\end{equation}

\subsection{Properties of correction functions from inverse powers}
\label{CorrProp}

\begin{figure}
\centerline{\includegraphics[width=10cm,keepaspectratio]{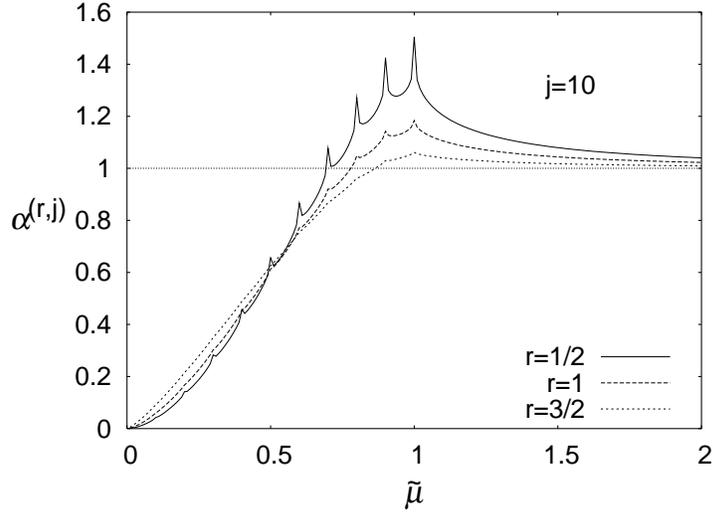}}
\caption{Behavior of the correction function $\alpha$ for larger $j$.
The general trend is similar to the case for $j=1/2$, but there are
$[j]+1$ spikes at $\mu=1,\ldots,j$ for integer $j$ and
$\mu=1/2,\ldots,j$ otherwise. Above the peak, the function is smooth.
\label{Fig_alpha10}}
\end{figure}

We have derived several different correction functions, making use
of different parameters $r$. In most cases one could make
different choices of such parameters and still write the
classically intended expression in an equivalent way. This gives
rise to quantization ambiguities since the eigenvalues of
$\hat{B}_{v,K}^{(r)}$ depend on the value $r$, and so will
correction functions.  In addition to the ambiguities in the
exponents $r$, one could use different representations for
holonomies before taking the trace rather than only the
fundamental representation understood above \cite{Gaul,Ambig}.  In
this case, we have more generally
\[
 \hat{B}_{v,K}^{(r,j)} =\f{3}{irj(j+1)(2j+1)}\left(2\pi\gamma
\lP^2\right)^{-1} \tr_j\left(
\tau^Kh_{v,K}\hat{V}_v^rh_{v,K}^{-1}\right)\,.
\]
Eigenvalues of such operators can be expressed as
\begin{equation} \label{BrjEigen}
 B_{v,K}^{(r,j)}=\f{3}{rj(j+1)(2j+1)}\left(2\pi\gamma
\lP^2\right)^{\f{3r}{2}-1} |\mu_{v,I}+\mu_{v,-I}|^{r/2}
|\mu_{v,J}+\mu_{v,-J}|^{r/2}
\sum_{m=-j}^jm|\mu_{v,K}+\mu_{v,-K}+2m|^{r/2}
\end{equation}
which leads to the general class of correction functions
\begin{equation} \label{alpharj}
 \alpha_{v,K}^{(r,j)}= \f{3}{r j
(j+1)(2j+1)}|\mu_{v,K}+\mu_{v,-K}|^{1-\f{r}{2}}
\sum_{m=-j}^j{m\left|\mu_{v,K}+\mu_{v,-K}+2m\right|^{r/2}} \,.
\end{equation}
After imposing isotropy the last expression becomes
\begin{equation}
 \alpha^{(r,j)}= \f{6}{r j (j+1)(2j+1)}|\mu|^{1-\f{r}{2}}\sum_{m=-j}^j{m\left|\mu+m\right|^{r/2}}\nonumber
\end{equation}
which is shown for a few cases in Fig.~\ref{Fig_alpha10}

For large $j$, the sum in $\alpha^{(r,j)}$ can be approximated using
calculations as in \cite{Ambig}. The idea is to consider two cases:
i) $\mu>j$ and ii) $\mu<j$ separately. In the former, the absolute
values can be omitted as all the expressions under the sum are
positive. Then the summation is to be replaced by integration to yield
\bq%
 \alpha^{(r,j)}= \f{12 j^3 \t\mu^{1-\f{r}{2}}}{r j
 (j+1)(2j+1)}&&\hspace{-1.5em}\left[\f{1}{r+4}\left((\t\mu+1)^{\f{r}{2}+2}-(\t\mu-1)^{\f{r}{2}+2}\right)\right.\nonumber\\
 &&\hspace{-1.5em}\left.-\f{\t\mu}{r+2}\left((\t\mu+1)^{\f{r}{2}+1}-(\t\mu-1)^{\f{r}{2}+1}\right)\right], \quad \t\mu>1 \nonumber
\eq%
where $\t\mu:=\mu/j$. In the second case, the terms in the sum
corresponding to $m<\mu$ and $m>\mu$ should again be considered
separately. The end result, however, is very similar to the
previous one
\bq%
 \alpha^{(r,j)}= \f{12 j^3 \t\mu^{1-\f{r}{2}}}{r j
 (j+1)(2j+1)}&&\hspace{-1.5em} \left[\f{1}{r+4}\left((\t\mu+1)^{\f{r}{2}+2}-(1-\t\mu)^{\f{r}{2}+2}\right)\right.\nonumber\\
 &&\hspace{-1.5em}\left.-\f{\t\mu}{r+2}\left((\t\mu+1)^{\f{r}{2}+1}+(1-\t\mu)^{\f{r}{2}+1}\right)\right], \quad \t\mu<1 \nonumber
\eq%
After some rearrangements and using that $j\gg 1$ these two
expressions can be combined into a single one as
\bq \label{alpharjapprox}
 \alpha^{(r,j)}= \f{6\t\mu^{1-\f{r}{2}}}{r
 (r+2)(r+4)}\left[(\t\mu+1)^{\f{r}{2}+1}(r+2-2\t\mu)+\sgn(\t\mu-1) |\t\mu-1|^{\f{r}{2}+1}(r+2+2\t\mu)\right]\,.
\eq%

\begin{figure}
\centerline{\includegraphics[width=10cm,keepaspectratio]{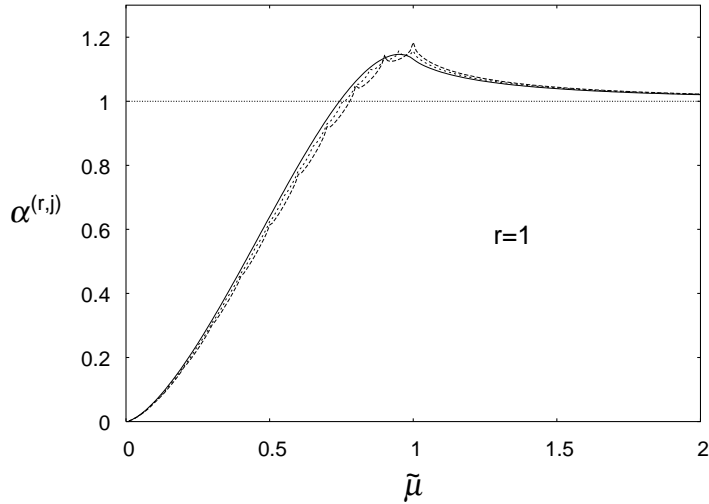}}
\caption{Comparison between the correction function (\ref{alpharj})
and its approximation (\ref{alpharjapprox}). The spikes are smeared
out by the approximation.
\label{Fig_alphacomp}}
\end{figure}

The approximation is compared to the exact expression of the
correction function obtained through eigenvalues in
Fig.~\ref{Fig_alphacomp}. As one can see, the spikes are smeared out
by the approximation (except for the point $\tilde{\mu}=1$ where the
approximation remains non-differentiable at second order which is not
visible from the plot). The general trend, however, is reproduced well
even below the peak. For applications in effective equations we note
that the approximation might be considered more realistic than the
exact eigenvalue expression because those equations would be based on
semiclassical states. Since such states cannot be eigenstates of the
triad but must only be peaked on a certain expectation value, they
will automatically give rise to a smearing-out of the spikes in the
eigenvalues as discussed in more detail in Sec.~\ref{ExpectVal}.

\subsubsection{Asymptotic behavior}

This class of correction functions parameterized by two ambiguity
parameters $r$ and $j$ captures the most important general properties
of such functions, including the position of their maxima at
$\t\mu\approx1$ (or $\mu\approx j$) and the initial power law increase
for small $\mu$ (determined by $r$) \cite{Ambig,ICGC}. It is indeed
easy to see that all correction functions have the correct classical
limit on large scales, such as 
\be\alpha^{(r,j)}(\t\mu)\approx 1+\f{1}{32
\t\mu^2}\f{(r-2)(r-4)}{3}\f{4(3j^2+3j-1)}{5}+O(\t\mu^{-4})
\rightarrow 1\ee for (\ref{alpha_iso}). Moreover, for small $\mu$
the correction function goes to zero as \be
\alpha^{(r,j)}(\t\mu)\approx (2\t\mu)^{2-\f{r}{2}},
\ee
which ensures boundedness of the quantized co-triad $e_{(K)}\propto
\alpha \sqrt{\t\mu} \propto \t\mu^2$ (using $r=1$ for this case as in
(\ref{cotriad})), when $\t\mu\rightarrow 0$. The same is true for
higher $j$ since evaluated at $\mu=0$ the sum of odd terms gives zero.

This function is not smooth but has a cusp at its maximum at
$\mu=1/2$, or more generally a cusp at integer or half-integer values
between $0$ and $j$. The second derivative $\alpha''$ is always
positive while $\alpha'$ changes sign between any two cusps. To the
right of the cusp at the largest $\mu$, the derivatives satisfy
\be\label{alpha_asymp}%
\alpha^\prime < 0, \quad \alpha^{\prime\prime} > 0 \,.
\ee
Note that the approximation used for larger $j$ smears out the cusps
and does not everywhere have positive second derivative. The behavior
above the peak and the general increase below is, however, reproduced
well by the approximation. The definite sign of $\alpha''$ has
far-reaching implications in the quantum corrected equations of motion
\cite{InhomEvolve}. Small corrections then add up during long cosmic
evolution times which would not be realized if, e.g., $\alpha$ would
oscillate around the classical value which is also conceivable a
priori.

\subsubsection{Small-scale behavior and ambiguities}

We will mostly use here and in cosmological applications of the
corrected perturbation equations of \cite{HamPerturb} the behavior for
larger values of $\mu$ above the peak. On very small scales, the
approach to zero at $\mu=0$ is special to operators with
U(1)-holonomies as they appear in the perturbative treatment here. In
particular, as we have seen explicitly the volume operator $\hat{V}$
and gauge covariant combinations of commutators such as $\tr(\tau^i
h[h^{-1},\hat{V}])$ commute. It is thus meaningful to speak of the
(eigen-)value of inverse volume on zero volume eigenstates. For
non-Abelian holonomies such as those for SU(2) in the full theory, the
operators become non-commuting \cite{DegFull}. The inverse volume at
zero volume eigenstates thus becomes unsharp and one can at most make
statements about expectation values rather than eigenvalues which
again requires more information on semiclassical states. Then, the
expectation values are not expected to become sharply zero at zero
volume, as calculations indeed show \cite{BoundFull}. In addition,
also here quantization ambiguities matter: We can write volume itself,
and not just inverse volume, through Poisson brackets such as
\cite{DegFull}
\begin{eqnarray*}
  V &=& \int\md^3x
\left(\frac{\epsilon^{abc}\epsilon_{ijk}}{6(10\pi
\gamma G/3)^{3}}
\int\md^3y_1\{A_a^i(x),|\det e(y_1)|^{5/6}\}\right.\\
&&\qquad\qquad\times\left.\int\md^3y_2\{A_b^j(x),|\det e(y_2)|^{5/6}\}
\int\md^3y_3\{A_c^j(x),|\det e(y_3)|^{5/6}\}\right)^2\,.
\end{eqnarray*}
After a lattice regularization as before, using
$V_v\approx\ell_0^3|\det(e_a^i)|$, we obtain
\begin{eqnarray*}
 V_v&=&\ell_0^6\left(\frac{|\det e|}{\sqrt{V_v}}\right)^2=
\ell_0^6\left(\epsilon^{abc}\epsilon_{ijk}
\frac{e_a^i}{V_v^{1/6}} \frac{e_b^j}{V_v^{1/6}}
\frac{e_c^k}{V_v^{1/6}}\right)^2\\
&=& \left(\frac{\epsilon^{abc}\epsilon_{ijk}}{6(10\pi
\gamma G/3)^{3}}\ell_0^3
\{A_a^i,V_v^{5/6}\} \{A_b^j,V_v^{5/6}\}
\{A_c^j,V_v^{5/6}\}\right)^2
\end{eqnarray*}
whose quantization, making use of commutators, differs from the
original volume operator (\ref{V_action}).  If non-Abelian holonomies
are used, it would not commute with the full volume operator of
\cite{AreaVol} or \cite{Vol2}. This clearly shows that the usual
quantization ambiguity also applies to what is considered the relevant
geometrical volume. (Related ambiguities for flux operators have been
discussed in \cite{Flux}.) It is not necessarily the original volume
operator constructed directly from fluxes, but could be any operator
having volume as the classical limit. For finding zero volume states
to be related to classical singularities, for instance, dynamics
indicates that volume constructed in the more complicated way through
commutators with the original volume operator is more relevant than
the volume operator constructed directly from fluxes \cite{DegFull}.
Thus, specific volume eigenstates have to be used with great care in
applications with non-Abelian holonomies. Also, the behavior of
correction functions below the peak value, especially whether or not
they approach zero at zero volume, is thus less clear in a general
context. In any case, below the peak positions scales are usually so
small, unless one uses larger $j$, that perturbation theory breaks
down. The behavior above the peak, by contrast, is robust and gives
characteristic modifications to the cosmological evolution of
structure.

\section{Effective Hamiltonian}
\label{EffHamDiscuss}

Calculations of distinct terms in the constraint presented in the
preceding section can now be used to derive effective Hamiltonian
constraints.

\subsection{Expectation values in semiclassical states and quantum
variables}
\label{ExpectVal}

The derivation of an effective Hamiltonian constraint proceeds by
computing expectation values of the constraint operator in
semiclassical states which are superpositions of our lattice states
peaked on perturbative metric and extrinsic curvature components.
Such states are easily constructible although, for the order we are
working at here, we do not need to do so explicitly. The peak values
of perturbative fields are thus in particular diagonal which means
that expectation values can easily be computed via Abelian
calculations.\footnote{Although initially SU(2)-holonomies appear in
the constraint, they only refer to lattice directions such that those
holonomies are of the form $\exp (A\tau_I)$. While these matrices do
not commute among themselves for different $I$, one can easily
re-arrange the order; see, e.g.,
\cite{CosmoI,SphSymm} for a discussion of the analogous effect in
symmetric models. The special form of SU(2)-matrices occurring in our
context is also the reason why we can take the trace in the
Hamiltonian constraint explicitly and reduce it to a product of sines
and cosines of curvature components.} The only complication arises
from the fact that we are necessarily dealing with operators as
products of holonomies and fluxes which are not simultaneously
diagonalizable. It is most convenient to use the triad eigenbasis
$|\ldots,\mu_{v,I},\ldots\rangle$ for triad or inverse triad
operators, and a holonomy eigenbasis for products of holonomies. This
was implicitly assumed previously in the curvature expansion and when
using inverse triad eigenvalues for correction functions.
However, for expectation values of the complete constraint operator as
a product of holonomy and co-triad terms we need to transform between
the two eigenbases, which as usually is possible by inserting
sums over complete sets of states: $\langle\psi|\hat{H}|\psi\rangle=
\sum_I\langle|\psi\hat{H}_1|I\rangle \langle I|\hat{H}_2|\psi\rangle$
if $\{|I\rangle\}$ is the complete set of states and
$\hat{H}=\hat{H}_1\hat{H}_2$ is factorized into the two parts
mentioned above. For a complete treatment we thus need to compute
matrix elements of $\hat{H}_1$ and $\hat{H}_2$, not just eigenvalues.

Nevertheless, the calculations presented before already provide the
main terms under the following approximation: We assume, without loss
of generality, that the complete set of states $\{|I\rangle\}$
contains a state $|\psi\rangle$ we are interested in. More crucially,
we assume that the spread $\sigma$ of $\psi$ in basic variables is
small. Under this assumption, $\langle I|\hat{H}_{i}|\psi\rangle$,
$i=1,2$, are dominated by $\langle\psi|\hat{H}_{i}|\psi\rangle$ since
(i) there is not much overlap with most other states in the complete
set and (ii) the states $|\psi\rangle$, having small spreads, are as
close as possible to eigenstates of $\hat{H}_1$ and $\hat{H}_2$,
respectively. With $\hat{H}_1$ being a product of holonomies and
$\hat{H}_2$ depending on fluxes, the spreads required in this
construction cannot be arbitrarily small because they are restricted
by uncertainty relations. This implies that additional corrections not
computed before arise due to the unavoidable spread of semiclassical
states. As a direct consequence of spreading, such terms depend on
parameters such as $\sigma$ which are nothing but the quantum
variables (\ref{QuantVars}) mentioned before. These variables
necessarily feature in a complete effective Hamiltonian, describing
how spreading and deformations of the state back-react on the peak
position \cite{Karpacz}.

Apart from these quantum variable terms, the main effective
Hamiltonian then is of the form $\langle\psi|\hat{H}_1|\psi\rangle
\langle\psi|\hat{H}_2|\psi\rangle$ where $|\psi\rangle$ is a
semiclassical state peaked on a given classical geometry. Each of the
two factors is of the form
\begin{equation}\label{convol}
 \sum_\mu O_{\mu}|\psi_{\mu}^{(\mu_0)}|^2\sim \sum_{\mu}
O_{\mu}f(\mu-\mu_0) \sim \int\md\mu O(\mu)f(\mu-\mu_0)=(O\star
f)(\mu_0)
\end{equation}
where $O_{\mu}$ are eigenvalues of an operator $\hat{O}$ and $\mu_0$
is the peak value of the state $|\psi\rangle$ in the basic variable
$\mu$ whose eigenbasis is used. On the right hand side, we see that
the effect of computing an expectation value in a semiclassical state
is mainly, to the given order, that eigenvalues appear in a form
convoluted with the shape of the semiclassical state.

In such a convolution, sharp features in eigenvalue functions such as
the spikes in Fig.~\ref{Fig_alpha10} will be smeared out. But
otherwise the general behavior is already displayed well by explicit
eigenvalues, and, similarly, higher order curvature corrections are close
to what we computed before. For general features we can thus avoid
dealing with details of states and their convolution with
eigenvalues. The form of effective Hamiltonians we arrive at is thus
\begin{eqnarray} \label{EffHam}
 H_{\rm eff}[N] &=& \frac{1}{8\pi G} \int_{\Sigma} \mathrm{d}^3x
N(x)\alpha[p]\left(-3\tilde{k}^2+
\Delta_K\right.\\
&&\qquad\qquad \left.
+\beta[p](-\tilde{p}^{-1}\partial^I\partial_I\tilde{p}+
\frac{1}{2}\tilde{p}^{-2}(\partial^I\tilde{p})(\partial_I\tilde{p}))+
\Delta^{(1)}_{\partial}\right)\sqrt{|\tilde{p}|} \nonumber\\
&&+
\int_{\Sigma}\md^3x N(x)\left(\frac{D[p]}{2|\tilde{p}|^{3/2}}\pi^2
+\frac{\sigma[p]|\tilde{p}|^{\frac{1}{2}}}{2}
(\partial^I\phi)(\partial_I\phi)
+\Delta^{(2)}_{\partial}
+|\tilde{p}|^{\frac{3}{2}}U(\phi)\right)\nonumber\\
&& +\int_{\Sigma}\md^3x N(x)\Delta_Q\nonumber
\end{eqnarray}
for metrics including scalar perturbations in longitudinal gauge. Note
that the correction functions depend functionally on the field $p(x)$,
not $\tilde{p}$, which shows that their scale is uniquely determined
by the state irrespective of choices of coordinates. Unspecified
correction terms are higher order curvature corrections $\Delta_K$
(see Eq.~(\ref{K_ave})), discretization corrections
$\Delta^{(1/2)}_{\partial}$ from different spatial derivative terms in
the constraint, and terms containing quantum variables $\Delta_Q$
which arise from metric as well as matter fields.

This form of effective constraints also demonstrates potential effects
of using SU(2) representations different from the fundamental
one. Notice that we did not compute this for the higher curvature
expansion since the required traces of different Pauli matrices are
more involved. But it is clear that this can only change the
coefficients in the expansion $\Delta_K$ since it always remains at a
perturbative level. Generally, larger values of $j$ mean that
curvature corrections will become important at smaller curvatures
compared to $j=1/2$, and thus coefficients in an expansion will
increase with $j$. The effect in inverse triad corrections
$\alpha[p]$, which we did compute explicitly here, is more pronounced
since $j$ determines the scale at which one enters the
non-perturbative regime of such inverse triad corrections. The main
difference between larger values of $j$ and the minimal one is that in
the former case peaked states exist whose spread is smaller than the
peak position of eigenvalues of an inverse triad operator. When this
is realized, the non-perturbative branch of increasing behavior
between $\mu=0$ and $\mu=j$ is not completely washed out in the
convolution but remains visible in an effective Hamiltonian.

In an effective Hamiltonian this consequence is obvious, but it was
not always clear from the underlying difference equation in isotropic
models. There, the discrete stepsize, determined by the
SU(2)-representation of holonomies in the constraint, equals the peak
position of inverse triad eigenvalues (see, e.g.,
\cite{AmbigConstr}). Thus, the discrete step in the difference
equation always jumps from zero directly to the peak when the same
representation is used for all holonomies occurring in the
gravitational as well as matter parts of the constraint. One could
thus argue that dynamics will be insensitive to the value of the
representation. Effective equations, if they are applicable in this
small-scale regime, show that this is not necessarily so. The
representation enters higher curvature terms differently from inverse
triad corrections, thus allowing effects of the non-perturbative part
to remain.

\subsection{Technical issues}

We now illustrate some of the more important choices we made in
constructing the constraint operators used here.

\subsubsection{Quantization procedures}
\label{QuantProc}

Our construction is suitable for a treatment of cosmological
perturbations within loop quantum gravity, but it does circumvent some
of its general aspects. First, we do not take into account full
non-abelian features; they can be included perturbatively but are not
required for our selection of mode and gauge. Secondly, we do not
allow irregular lattices or valence higher than six. Also this can be
included by summing operator contributions over triples of edges (as
they are constructed in the full setting). Detailed coefficients in
effective expressions may then change, but not qualitative
effects. Moreover, as already mentioned the labels coming with
additional edges or higher valent vertices are redundant for
cosmological perturbations.

We have presented higher power corrections using ``holonomies'' based
on extrinsic curvature rather than connection components since this
simplifies the calculations considerably. Using the
background, it is mathematically possible to define such objects,
although in a full background independent setting only holonomies of
a connection would be well-defined. We use this mainly as a first
possibility to demonstrate which types of corrections one expects and
will discuss now how general the resulting expressions
can be considered to be. This refers to corrections to terms of the
Lorentzian constraint which, schematically, can be written as
\begin{equation}\label{FK}
 F+K^2=\md A+A^2+(A-\Gamma)^2= \md(\Gamma+K)+(\Gamma+K)^2+K^2
\end{equation}
to be multiplied by triad components dealt with by Poisson brackets.

Using extrinsic curvature as basic object, one obtains trigonometric
functions of its components which when expanded give higher power
corrections to $\md K+K^2$. But since the spin connection has been
split off from the basic object, one has to quantize it individually
and add suitable combinations for $\md\Gamma+\Gamma^2$ to the
constraint. Here, we assume that the cross-term $\Gamma K$ does not
contribute which is indeed the case for diagonal triads (implying
anti-symmetric spin connection components) and extrinsic
curvature. This is not much of a restriction: $K$ is required to be
diagonal for $K$-holonomies to simplify the calculations. Moreover,
the perturbative non-diagonal part of $\Gamma$ must be antisymmetric
because it perturbs an SO(3)-matrix. If there is a diagonal
contribution, e.g.\ from a spatially curved background, it can be
combined with $K$. As for corrections, we have higher power
corrections in the quantization of $\md K+K^2$ and inverse triad
corrections in $\md\Gamma+\Gamma^2$ since the spin connection contains
inverse triad components.

Using $A$-holonomies gives, at first sight, a different picture. Now,
$F=\md A+A^2$ receives higher power corrections, but the spin
connection is not quantized directly, not giving immediate inverse
triad corrections. One rather has to proceed as in the full theory
\cite{QSDI} where the term $(A-\Gamma)^2=K^2$ in the
constraint is re-written using (\ref{Kcomm}). The double-Poisson
bracket (\ref{Kcomm}) used to quantize extrinsic curvature now leads
to additional corrections. In particular, since inverse triad
quantizations have been used in $H^{(1)}$ in (\ref{Hone}),
corresponding corrections do arise which are qualitatively similar to
those in a direct quantization of the spin connection. One thus
expects similar types of corrections, as with $K$-holonomies and
$\Gamma$-quantizations, although in different combinations.

In our construction, $K$-holonomies arose from $A$-holonomies through
a perturbative expansion in non-diagonal components.  When using
$A$-holonomies on a spatially curved background such as a closed
Friedmann--Robertson--Walker model, it is furthermore necessary to
take lattice effect of an inhomogeneous model
\cite{InhomLattice} or related effects \cite{APS} into account. A
fine enough lattice (${\cal N}\gg1$) is required for a semiclassical
expansion of holonomies since
\[
 \exp(iA)\sim\exp(i(\Gamma+K))\sim
 \exp(i(K+{\cal N}^{-1/3}\bar{\Gamma}+\delta\Gamma))
\]
with the background spin connection possibly of the order
$\bar{\Gamma}=V_0^{1/3}\tilde{\bar{\Gamma}}=O(1)$ can be expanded in
all terms only if ${\cal N}$ is large. (For a closed isotropic model, for
instance, $\bar{\Gamma}=1/2$
\cite{Closed}.) The number of vertices ${\cal N}$ enters through
$\ell_0\tilde{\bar{\Gamma}}={\cal N}^{-1/3}\bar{\Gamma}$ in holonomies. The
spin connection perturbation $\delta\Gamma$ will be small in
perturbative regimes such that we can always write
\[
 \exp(i(K+{\cal N}^{-1/3}\bar{\Gamma}+\delta\Gamma))=
\exp(i(K+{\cal N}^{-1/3}\bar{\Gamma}))(1+i\delta\Gamma+\cdots)\,.
\]
But the remaining exponential also has to reduce to the leading terms
of an expansion in semiclassical regimes. While $K$ is then
automatically small, this may not be the case for $\bar{\Gamma}$.
Without the reduction by ${\cal N}^{-1/3}$ for a fine lattice, one could not
expand the exponentials to reproduce the polynomials in $K$ and
$\Gamma$ classically occurring in the constraints.

We have focused here on the first type of quantization which is
simpler to compute explicitly but may not be as close to the full
theory. Using $A$-holonomies, no diagonality can be used, but
perturbative treatments of non-diagonal components are possible. It is
thus possible, though more involved, to compute correction terms
obtained through different quantization schemes and to compare their
consequences, in particular those at the phenomenological level. A
first result in this direction follows from the perturbation equations
derived in \cite{HamPerturb} which show that effects of inverse triad
corrections in $\Gamma$ are less significant than those in the
commutator \cite{InhomEvolve}. Thus, one can hope that the precise
quantization procedure of curvature is not very important for physical
aspects accessible so far. A detailed investigation of all
consequences can nevertheless provide important guidance as to which
procedure should be pursued in the full theory.

\subsubsection{Different types of correction functions}

In fact, we have included in the computation of perturbative effects
four different correction functions $\alpha$, $\beta$, $D$ and
$\sigma$. All of them come from inverse triad corrections. With all
functions coming from the same type of modification, one may wonder
why they should not all be identical.

It is clear from the procedure that these functions arise from
different classical functionals of densitized triad components. For
instance, $\alpha$ comes from the antisymmetric part of
$E^a_iE^b_i/\sqrt{|\det E|}$ while $\sigma$ comes from the symmetric
part. They could be related to the same correction, but the
quantization requires quite different rewritings (\ref{AB_def}) and
(\ref{gradient}) of the corresponding terms in Hamiltonians such that
correction functions will differ. In particular, they come with
different parameters such as $r$. On top of that, each correction
function is subject to quantization ambiguities. As we have seen,
however, the typical behavior is robust under changes of the
parameters. In particular, all correction functions have the same
qualitative properties and differ only quantitatively in a way
parameterized by a few parameters.

\subsubsection{Implications for gauge issues}

The assumptions on states used to derive effective constraints have a
bearing on the gauge issue. By specifying the peak value of a spatial
geometry and its extrinsic curvature in a semiclassical state we are
fixing the spatial diffeomorphism constraint rather than solving it by
averaging as done in the full theory \cite{ALMMT}. Choosing the form
of peak values partially implements a chosen gauge, but still allows
some freedom. We also note that even though spatial diffeomorphisms
are fixed, one still has to impose the constraint. This will give rise
to one of the cosmological perturbation equations as is clear from
\cite{HamPerturb}.

Fixing the diffeomorphism constraint also implies a different
viewpoint for the Hamiltonian constraint operator of the loop
quantization. In the full construction \cite{QSDI}, one makes use of
diffeomorphisms in order to make the operator more independent of the
choice of edges used to quantize curvature. When diffeomorphisms are
fixed, this is no longer possible and effective constraints would
depend on precisely how such edges are chosen. We have fixed this
freedom here by laying the edges entirely on the lattice resulting in
a graph preserving operator. Thus, holonomy corrections in the
constraint depend on the lattice spacing provided through a state
implementing the background geometry. While this simplifies the
calculations without leading to significant quantitative changes in
coefficients, we are as a consequence disregarding the creation of new
vertices by the constraint operator. Thus, ${\cal N}$ is constant for
the construction, but may effectively be assumed to be slowly
dependent on, e.g., the total volume (see \cite{InhomLattice} for more
details).

\section{General implications for effective theory}
\label{Effective}

Our calculations, following the scheme to derive effective equations
sketched before, have led to corrections (\ref{EffHam}) which arise as
leading order terms in an effective theory of perturbative loop
quantum gravity. No complete expression has been derived, but
characteristic terms are clear and lead to interesting phenomena
\cite{InhomEvolve}. Rather than studying one model in detail we have
provided here an illustration of the general scheme: The
characteristic feature of loop quantizations is the use of holonomies,
which give rise to typical correction terms. They can be split into
higher power corrections, which are always perturbative, and
corrections to inverse powers of triad components which become
non-perturbative at small scales. All these corrections are in
addition to discretization and genuine quantum effects such as higher
time derivatives. In this section we highlight conceptual conclusions
that can be drawn from such a scheme.

\subsection{Basic variables in quantum gravity corrections}

Holonomy corrections arise through expectation values and thus depend
on the basic variables used in the quantization.  Using commutators to
quantize triad components, for instance, modifies the classical
expressions in a way which can be computed through the explicitly
available eigenvalues such as (\ref{BrjEigen}) of these operators. The
occurrence of trigonometric functions instead of direct curvature or
connection components leads to higher power terms when expanded in an
effective constraint. Such corrections depend, by construction, on
$p(x)/\lP^2=\ell_0^2\tilde{p}(x)/\lP^2$ and $k(x)=\ell_0\tilde{k}(x)$,
respectively, both of which are independent under rescaling the
coordinates. They do depend, however, on the lattice size which
determines the scales on which a state probes the field.

If we split the fields into background parts
\begin{equation}
 \tilde{\bar{p}}:=\frac{1}{V_0} \int\md^3x
\tilde{p}(x)\quad\mbox{and}\quad
\tilde{\bar{k}}:=\frac{1}{V_0}\int\md^3x\tilde{k}(x)
\end{equation}
by integrating over a cube (sufficiently large to contain, say, the
Hubble volume) of coordinate volume $V_0=\int\md^3x$ and perturbations
\begin{equation}
 \delta \tilde{p}(x)= \tilde{p}(x)-\tilde{\bar{p}}\quad\mbox{and}\quad
\delta\tilde{k}(x)= \tilde{k}(x)-\tilde{\bar{k}}
\end{equation}
to set up cosmological perturbation theory
\cite{HamPerturb},\footnote{In \cite{HamPerturb}, only written here
with a tilde have been used, but the tilde was dropped for notational
convenience.} we can see from preceding constructions that it is not
these fields directly which occur in correction functions. In
isotropic loop quantum cosmology, the quantization is based on
variables
\begin{equation}
 \bar{p}=V_0^{2/3}\tilde{\bar{p}}\approx\frac{1}{V_0^{1/3}}\sum_v
\ell_0^3\tilde{p}(v)=\frac{1}{{\cal N}^{1/3}}\sum_vp(v)
\end{equation}
and
\begin{equation}
 \bar{k}=V_0^{1/3}\tilde{\bar{k}}\approx\frac{1}{V_0^{2/3}}\sum_v
\ell_0^3\tilde{k}(v)=\frac{1}{{\cal N}^{2/3}}\sum_vk(v)
\end{equation}
which now appear as lattice averages in an inhomogeneous setting and
provide the background for cosmological perturbation theory. As such,
they do not depend on the auxiliary coordinate volume $V_0$ as they
would in homogeneous models \cite{Bohr} but on the number ${\cal N}$
of lattice vertices. These two quantities are related by $V_0={\cal
N}\ell_0^3$ through the lattice size $\ell_0$, but ${\cal N}$ has
significance as a parameter specifying the states rather than just
being auxiliary as $V_0$. Similarly, basic variables of the
inhomogeneous theory are functions\footnote{Note that $\ell_0$ is used
to rescale the inhomogeneous $\tilde{p}(x)$ while $V_0$ is used to
rescale $\tilde{\bar{p}}$ as it is done in isotropic models
\cite{Bohr}.}
\begin{equation} \label{psplit}
 p(x)=\ell_0^2\tilde{p}(x) =\frac{\ell_0^2}{V_0^{2/3}}\bar{p}+\delta
p(x)= \frac{1}{{\cal N}}\sum_vp(v)+\delta p(x)
\end{equation}
which directly occur in correction functions through fluxes, and
\begin{equation}
 k(x)=\ell_0\tilde{k}(x) =\frac{\ell_0}{V_0^{1/3}}\bar{k}+\delta
k(x)= \frac{1}{{\cal N}}\sum_vk(v)+\delta k(x)
\end{equation}
occurring in higher power corrections through holonomies.

This shows that the resulting equations are rescaling invariant when
$\tilde{p}$, $\tilde{k}$ and $\ell_0$ change simultaneously, a fact
which was not always obvious in isotropic models based on the scale
factor. As expected, the equations are also dependent on specifics
(mainly ${\cal N}$) of the state whose dynamics is described
effectively. This shows which states are suitable for perturbation
theory and when perturbations break down. A perturbation scheme works
only if $\delta \tilde{p}\ll\tilde{\bar{p}}$ which from (\ref{psplit})
implies that differences between local edge labels of the state
(corresponding to $\delta p(x)=\ell_0^2\delta\tilde{p}(x)$) must be
small compared to the average lattice label ${\cal N}^{-1}\sum_vp(v)$
(corresponding to the perturbative background value of $p(x)$). Since
the labels are discrete, differences between them have a positive
lower bound unless they are equal. Thus, the average label must be
large compared to the discrete gap in the spectrum of labels. In our
U(1)-theory, labels are integer valued which means that the average
label must be larger than one, and local edge labels must not stray too
much from the average. There is no such restriction from the curvature
perturbations because curvature does not have a discrete spectrum.

\subsection{Quantum variables and classical limit}

Starting from the Hamiltonian (constraint) operator in any quantum
theory, the quantum Hamiltonian is defined as a function on the
projective Hilbert space determined by taking expectation values. This
can be seen as the Hamiltonian function of a dynamical system whose
phase space is obtained from the Hilbert space
\cite{GeomQuantMech,ClassQuantMech,Schilling}. The system thus appears of
classical form at least as far as dynamics is concerned, but each of
its classical degrees of freedom is accompanied by infinitely many
quantum variables (\ref{QuantVars}). An effective description requires
a further step, truncating the infinitely many quantum variables to a
finite set \cite{EffAc}. If this is done consistently, one obtains
effective equations which amend the classical ones by quantum
corrections. One often performs such a truncation by using a certain
class of semiclassical states to compute expectation values of the
Hamiltonian operator. The regime under consideration determines what a
suitable set of semiclassical states is.

Based on the assumed semiclassicality of states peaked at values $p^I$
and $k_J$, the expressions we derived give the main part of the
effective Hamiltonian constraint computed as an expectation value in
such states. Note that we did not explicitly compute expectation
values in states but read off corrections from operators by expanding
trigonometric functions arising from holonomies or eigenvalues of
inverse triad operators. Each of these corrections requires, strictly
speaking, eigenstates of holonomies for higher order corrections in
$\tilde{k}$ or flux eigenstates for corrections as functions of
$\tilde{p}$. But even if we were to compute expectation values in
peaked states, the main corrections would be of the form read off from
different eigenstates as seen by analogous calculations in the
isotropic case \cite{Bohr,Josh}. In general, one has to use
semiclassical states which are neither eigenstates of holonomies nor
of flux operators. This gives rise to additional contributions
depending, e.g., on the spread of the state.  From the spread and
other detailed properties of states one obtains contributions
depending on additional independent quantum variables corresponding to
fluctuations and correlations.  For non-quadratic Hamiltonians or
constraints, these quantum variables couple to classical variables and
influence their motion. To some degree, the appearance of additional
independent quantum variables corresponds to higher derivative terms
in effective actions \cite{Karpacz}. Thus, we obtain modified
coefficients (from correction functions such as $\alpha$), higher
powers in momenta (from $\sin k_I$ and $\cos k_I$) and higher
derivative terms from quantum variables (interpreted as higher time
derivatives) and from the discretization (higher spatial derivatives),
which comprise all effects known and expected from effective
actions. The first two arise as typical corrections by using
holonomies.

\subsubsection{Basic variables vs.\ coarse graining}
\label{Coarse}

In our treatment here we assumed that lattice scales are small
compared to other scales of the relevant physical fields such as
matter or classical metric modes to be obtained in a semiclassical
limit. In such a context, it is sufficient to use the basic variables
as they come as labels of a quantum state directly in effective
correction functions. This is not possible in regimes where basic
variables of the states are themselves strongly inhomogeneous as it
necessarily happens when the discrete flux labels $\mu_{v,I}$ approach
the lowest non-zero value one. Then, the perturbative condition
$\delta\mu_{v,I}\ll\mu_{v,I}$ where $\delta\mu_{v,I}$ refers to the
difference in nearby labels cannot be satisfied unless
$\delta\mu_{v,I}=0$, i.e.\ the labels are exactly homogeneous. Most
likely, this happens in strong curvature regimes where perturbation
theory would be expected to break down even classically. But since the
discreteness of the labels $\mu_{v,I}$ plays a role in this simple
argument, there can be regimes where classical perturbation theory
would be applicable but the underlying lattice formulation would not
seem to be in a perturbative regime. In such cases, one would have to
coarse grain the basic variable, i.e.\ replace the basic lattice site
variables by averages over larger patches of an intermediate
scale. Then, the averaged labels would increase, relieving the
contradiction between $\delta\mu_{v,I}\ll\mu_{v,I}$ and quantum
discreteness.

\subsubsection{Orders of magnitude of corrections}
\label{magnitude}

With several different correction terms, it is helpful to know whether
in certain regimes some of them can be ignored. This can be difficult
to determine in homogeneous models unless one makes special choices of
ambiguity parameters such as large values of $j$ \cite{Inflation}. In
inhomogeneous situations it is often simpler to determine which
corrections are expected to be dominant because they depend
differently on the basic scale contained in $p_{v,I}$
\cite{InhomLattice}. These variables are parameters determining the
state and thus the physical regime being probed. When $p_{v,I}$ is
small, i.e.\ close to its minimum $\ell_{\rm P}^2$, inverse triad
corrections are large. They decrease when $p_{v,I}$ becomes larger,
but this also implies larger and fewer lattice sites such that
discretization effects become important. Moreover, in nearly isotropic
configurations extrinsic curvature is given by
\begin{equation} \label{kest}
 k_{v,I}=\sqrt{8\pi Gp_{v,I}\rho/3}
\end{equation}
as it
follows from the Friedmann equation. The energy density scale $\rho$
thus determines when curvature corrections are relevant. Since there
is also a factor of $p_{v,I}$, curvature corrections increase with
larger $p_{v,I}$ just as discretization corrections.

For a semiclassical regime we must have $p_{v,I}>\ell_{\rm P}^2$ in
order to reproduce closely the correct inverse powers of triad
components. We must also have a discreteness scale $p_{v,I}$ which is
sufficiently small in order to avoid discretization effects already
in, say, particle physics. This requires $p_{v,I}$ to be much smaller
than the typical physical scale squared, such as a wave length
$\lambda$ of field modes or even the Hubble length
$a/\dot{a}$. We thus have a range $\ell_{\rm P}^2<p_{v,I}\ll
\lambda^2$ or $\ell_{\rm P}^2<p_{v,I}\ll (8\pi G\rho)^{-1}$ if we
express the Hubble length in terms of energy density. At the upper
bound we ensure that discretization effects do not disrupt other
physics used essentially in a given scenario. As a consequence of
(\ref{kest}), this {\em implies} that higher order corrections in
curvature are small, too. The dominant contributions are then given by
inverse triad corrections which we have focused on in the preceding
derivations. Note that the semiclassical range for $p_{v,I}$ is large
in late time cosmology, implying that corrections can be arbitarily
small, for instance to the propagation of signals from gamma ray
bursts. In the early universe, however, and in particular during
inflation the energy scale is much higher, restricting the range more
narrowly \cite{InhomEvolve}. The best tests of quantum gravity effects
are thus expected from early universe cosmology.

\subsubsection{Classical limit}

We have ignored in our calculations so far any detailed specifics of
states and terms containing quantum variables. Implicitly, we are thus
assuming that such terms are subdominant, just as one assumes
analogous terms to be subdominant in a derivative expansion of low
energy effective actions. Under this assumption we reproduce classical
expressions in the suitable limit, which proves that loop quantum
gravity has the correct classical limit in this perturbative regime in
the same sense as in usual effective theories. This statement
certainly includes inhomogeneities in the perturbative sector
considered here. For instance, the Newton potential and corrections on
smaller scales can be obtained from perturbation equations derived
from the effective constraints \cite{InhomEvolve}.

In effective theory, verifying the correct classical limit does not
require one to construct explicit dynamical coherent states, not even
approximately. This would certainly be of interest, but would be
highly complicated and is rarely done in interacting field theories
where one can nevertheless be certain about the correct classical
limit. We emphasize that, in any case, a discussion of the
semiclassical limit based on coherent states does require such states
to be {\em dynamical} coherent states, or at least must involve
statements on dynamical changes of state parameters to specify
suitable regimes. This means that states must stay approximately
coherent under evolution or, in a fully constrained theory such as
gravity, solve the Hamiltonian constraint. If this is not realized,
quantum variables and the back-reaction of spread and deformations on
the classical variables are not under sufficient control to ensure the
correct classical limit.

There are two viable procedures to verify the correct classical limit
of a quantum theory, be it a constrained or unconstrained
system. First, one may use kinematical coherent states to compute
expectation values of the dynamical operators (a Hamiltonian or
constraint operators) and then analyze the dependence of quantum
variables in resulting equations of motion; if their effect on
expectation values is small in suitable regimes, the correct classical
limit results. Secondly, dynamical coherent states can be used if they
can be constructed at least approximately, which directly illustrates
whether the dynamics of expectation values is close to the classical
one. The second procedure is much more complicated for interacting
theories since the full quantum dynamics would have to be solved at
least approximately at the quantum level. The first procedure, by
contrast, allows one to derive effective dynamical equations first and
then approximate solutions to understand the behavior of expectation
values. Thus, usually kinematical semiclassical states are used in
explicit effective descriptions, followed by an analysis such as one
in a derivative expansion in quantum field theory. Such a further
analysis is always required when kinematical semiclassical states are
used, and it can only be done in a regime dependent way to bring in
conditions for when semiclassicality should be satisfied. We have done
this implicitly in our discussion by assuming slowly varying fields as
in usual derivative expansions, both in space by doing a continuum
limit of the lattice states and in time by assuming quantum variables
to be negligible.

We emphasize again that even if one can demonstrate an ``instant''
classical limit by using kinematical coherent states, a dynamical
statement would still require one to assume (or to show) that such
back-reaction effects of quantum variables on expectation values are
not strong. This picks the correct regime of states in which one has
semiclassical behavior. Without such an additional analysis,
kinematical coherent states would neglect the back-reaction of
spreading and deformations of states on expectation values which are
essential for dynamical effective equations \cite{EffAc}.  An
additional aspect arises for generally covariant situations where not
all variables can be peaked in a semiclassical state as it would be
the case in a common kinematical coherent state. Some of the phase
space variables will have to play the role of internal ``clocks'' in
which evolution of expectation values as well as quantum variables is
measured. Thus, when constructing kinematical coherent states to check
the classical limit, they must not be peaked on all phase space
variables; a choice of clock has to be made before the
calculation. Then, quantum variables also back-react on the change of
the clock.

Often, investigations of classical limits based on kinematical
coherent states are motivated by well-known constructions of the
harmonic oscillator or free quantum field theories. The behavior of
quantum variables or of dynamical coherent states in general can,
however, be very different from the well-studied aspects of the free
systems. Such systems or small deviations from them with anharmonicity
can well be studied by coherent state techniques. But gravity is very
different and not expanded around a set of harmonic oscillators. In
fact, gravity with its unbounded Hamiltonian even lacks a ground state
or vacuum to expand around. The bounce model solved in
\cite{BouncePert}, for instance, shows that the spreads change
exponentially rather than being constant or at least periodic as it
happens for the harmonic oscillator. The resulting semiclassical
picture is very different from that provided by harmonic oscillator
coherent states. This must be taken into account in semiclassical
analyses; effective theory provides suitable means to study such
situations in sufficiently general terms as initiated in this paper.

\subsection{Collective graviton}

The constructions indicate a picture of the classical limit of quantum
gravity where linear metric modes appearing in the evolution equations
are not basic excitations of a quantum field. They arise, rather, as
collective excitations out of the underlying discrete quantum
theory. At a basic level, degrees of freedom are encoded in quantum
numbers $\mu_{v,I}$ while the scalar mode, for instance, is obtained
through the difference between such a local label and the average
value on the whole lattice. The classical modes thus arise as
non-local, effective excitations out of the underlying quantum state
\cite{InhomLattice}. This shows in a well-defined sense how classical
degrees of freedom are obtained as collective excitations, analogously
to phonons in a crystal. That the correct classical dynamics results
for these collective modes is demonstrated, for instance, by the
derivation of Newton's potential for perturbations on a flat isotropic
background in \cite{InhomEvolve}.

\section{Summary}

Together with \cite{EffAc,InhomLattice,HamPerturb} we have shown in
this paper that techniques are now available to derive effective
equations of cosmological perturbation theory. The geometrical
background on which cosmological perturbations are defined is
introduced through a class of states, rather than being used to set up
the quantization. Background independent quantum properties thus
remain, but one can make use of perturbation expansions for explicit
calculations. The role of quantum labels and basic variables is clear
from this procedure, which determines the type as well as order of
magnitudes of correction terms which remained obscure previously. The
inhomogeneous treatment including all relevant modes allows us to see
all possible correction terms. Note, for instance, that since the
isotropic expression for the co-triad is finite on small scales even
classically, it would not contribute a correction function to the
gravitational part of the Hamiltonian constraint in a purely isotropic
setting. When isotropic expressions are quantized directly, a
cancellation of the inverse isotropic triad component hides possible
quantum effects of the full constraint. This does not arise in our
context starting from an inhomogeneous lattice quantization. Thus,
complete corrections are obtained in reliable form.

Many different regimes are still to be explored to obtain a full
overview of all effects. Moreover, gauge issues have to be
investigated which is of relevance for the full quantum theory,
too. While general effective equations including the relevant
inhomogeneities are now available and orders of correction terms can
be estimated, one still has to use them with care since they are not
yet formulated for gauge invariant perturbations. In this context one
should notice that not only evolution equations but also gauge
transformations are determined by the constraints and thus modified by
quantum corrections. It is thus not possible to use classical
expressions for gauge invariant quantities since they will receive
additional corrections. These issues are currently being studied to
complete the derivation of equations with quantum corrections. The
strategy for computing those corrections from quantum operators has
been provided in this paper.

When speaking of quantum corrections to classical equations it is
clear that to zeroth order the correct classical limit has to be
satisfied. In fact, what we have shown here implies that loop quantum
gravity has the correct classical limit for scalar modes in
longitudinal gauge propagating on a spatially flat background. In the
process, we have demonstrated which steps must be involved in such a
detailed calculation, most importantly a continuum limit but also a
slowly varying field approximation. Extensions to other modes and
gauges, and different backgrounds, can be done by the same techniques
but are technically more involved to do explicitly. Nevertheless, it
is clear from the construction that the correct classical limits will
also be reproduced in those cases. More precisely, we have shown in
Sec.~\ref{magnitude} that there are always ranges of the basic lattice
variables such that quantum corrections are small in nearly classical
situations of low energy density and small curvature. In more
energetic cosmological situations, those ranges can shrink to narrow
intervals such that significant quantum corrections can be expected
\cite{InhomEvolve}.

This demonstration of the correct classical limit crucially rests on a
new understanding of effective theory \cite{EffAc}. Although it has
not yet been formulated fully for field theories (but see
\cite{EffectiveEOM}), this scheme is applicable here due to the
ultraviolet cut-off of quantum gravity. On any lattice state we have
only finitely many degrees of freedom in any compact spatial volume to
which the quantum mechanical techniques of \cite{EffAc} directly
apply. While loop quantum gravity does not possess a sharp cut-off but
is rather based on arbitrary graphs in space any of which can occur in
a general superposition \cite{ALMMT}, effective equations are always
defined with respect to a single class of states. The physical state
thus determines the cut-off dynamically. We have certainly not used
explicit physical solutions of the Hamiltonian constraint but rather
computed effective equations from general states. If a physical
solution were available, the labels $p_v$ would be determined
explicitly and fix the order of correction terms completely. Moreover,
a full solution would determine how the lattice itself changes by the
creation of new vertices in terms of an internal clock such as the
total volume. This graph-changing nature seems to be one of the most
important effects to be understood especially for late-time evolution
in cosmology, or any dynamical issue relevant for large spatial
slices.  Although such a full solution seems currently out of reach,
models and effective analyses already provide quite detailed
information on the dynamics of background independent quantum gravity
in cosmologically relevant regimes.

\section*{Acknowledgements}

MB was supported by NSF grant PHY-0554771, HH by the fellowship
A/04/21572 of Deutscher Akademischer Austauschdienst (DAAD) and MK by
the Center for Gravitational Wave Physics under NSF grant
PHY-01-14375. We thank Parampreet Singh for joining us in initial
calculations of this project.


\end{document}